\documentclass[3p,12pt]{elsarticle} 

\usepackage{amsfonts,amsmath,amssymb,graphicx}
\usepackage{color,mathrsfs,bm,appendix,subfig}
\usepackage[dvipsnames]{xcolor}

\usepackage[figuresright]{rotating}
\usepackage{multirow}
\usepackage{colortbl}

\usepackage{natbib}
\usepackage{hyperref}

\usepackage{epstopdf}
\usepackage{capt-of}
\usepackage{lineno}
\usepackage{booktabs}

\journal{XXX}

\begin{document}

\title{A fast spectral method for the Uehling-Uhlenbeck equation for quantum gas mixtures: homogeneous relaxation and transport coefficients}

\author[strath]{Lei Wu~\corref{cor1}}


\cortext[cor1]{Corresponding author.\\
	E-mail address: lei.wu.100@strath.ac.uk (L. Wu). }	

\address[strath]{James Weir Fluids Laboratory, Department of Mechanical and Aerospace Engineering, University of Strathclyde, Glasgow G1 1XJ, UK }


\begin{abstract}

A fast spectral method (FSM) is developed to solve the Uehling-Uhlenbeck equation for quantum gas mixtures with generalized differential cross-sections. The computational cost of the proposed FSM is $O(M^{d_v-1}N^{d_v+1}\log{N})$, where $d_v$ is the dimension of the problem, $M^{d_v-1}$ is the number of discrete solid angles, and $N$ is the number of frequency nodes in each direction. Spatially-homogeneous relaxation problems are used to demonstrate that the FSM conserves mass and momentum/energy to the machine and spectral accuracy, respectively. Based on the variational principle, transport coefficients such as the shear viscosity, thermal conductivity, and diffusion are calculated by the FSM, which compare well with analytical solutions. Then, we apply the FSM to find the accurate transport coefficients through an iterative scheme for the linearized quantum Boltzmann equation. The shear viscosity and thermal conductivity of the three-dimensional quantum Fermi and Bose gases interacting through hard-sphere potential are calculated. For Fermi gas, the relative difference between the accurate and variational transport coefficients increases with the fugacity; for Bose gas, the relative difference in the thermal conductivity has similar behavior as the gas moves from the classical to the degenerate limits, but that in the shear viscosity decreases. Finally, the shear viscosity and diffusion coefficient have also been calculated for a two-dimensional equal-mole mixture of Fermi gases. When the molecular mass of the two components are the same, our numerical results agree well with the variational solution. However, when the molecular mass ratio is not one, large discrepancies between the accurate and variational results are observed; our results are reliable because (i) the method relies on no assumption and (ii) the ratio between shear viscosity and entropy density satisfies the minimum bound predicted by the string theory. 


\end{abstract}


\maketitle

\section{Introduction}

The experimental manipulation of ultracold atomic gases has attracted extensive research interest to understand the dynamic of quantum systems~\cite{RMP1}. Most researches focus on the condensed phases~\cite{Anderson1995Science,Jin2003Nature}, since these quantum systems are ideal to study the crossover from a Bardeen-Cooper-Schrieffer superfluid to Bose-Einstein condensation, which is ubiquitous in high-temperature superconductivity, neutron stars, nuclear matter, and quark-gluon plasma. In experiments, however, they are prepared from dilute gases at room temperature, where the thermal motion of gas molecules is described by the Boltzmann equation. As the temperature goes down, the thermal de Broglie wavelength could become comparable to the interatomic distance; in this case the quantum effects emerge, and the thermal motion of quantum gases can be described by the Uehling-Uhlenbeck equation~\cite{Uehling1933}, which is also known as the quantum Boltzmann equation (QBE). When the temperature decreases further, the condensation begins, and the condensed phase coexists with the normal phase. For example, for Bose gas, at the temperature below the onset of the Bose-Einstein condensation, the QBE and Gross-Pitaevskii equation are used to describe the dynamics of the Bose gas in the normal and condensed phases, respectively~\cite{JacksonPRL2001,JacksonPRA2002}; the exchange of gas molecules between the normal and condensed phases is also described by the Boltzmann-type collision operators.

Mathematically speaking, the QBE, which is defined in the six-dimensional phase space, is much more complicated than the Gross-Pitaevskii equation in the three-dimensional physical space. Although in the hydrodynamic regime (i.e. when the mean free path of gas molecules and the characteristic oscillation frequency are respectively much smaller than the characteristic flow length and the mean collision frequency of quantum gases) the Navier-Stokes equation can be derived from the QBE via the Chapman-Enskog expansion~\cite{CE} to describe the gas dynamics, in quantum experiments, however, this situation is always violated: since the gas is confined by external potentials, the gas density is very small in the vicinity of the trap so that its dynamics is highly rarefied. Therefore, to describe the dynamics of quantum gas in the normal phase accurately, an efficient and accurate method to solved the QBE is necessary. In the paper we focus on the QBE only.

The direct simulation Monte Carlo method (DSMC)~\cite{JacksonPRL2001,JacksonPRA2002,GarciaPRE2003} has been proposed to solve the QBE. Since the collision frequency is enhanced (or reduced) for Bose (or Fermi) gas, and this enhancement (or reduction) relies on the velocity distribution function (VDF) after the binary collision, the DSMC method for QBE needs to use a very large number of simulated particles to sample the VDFs, which is in sharp contrast to the DSMC for classical gases where no such sampling is needed~\cite{Bird1994}; for Fermi gas, due to the Pauli exclusion principle, the collision frequency might even become negative (unphysical) if the VDF is not accurately sampled~\cite{BorowikJCP2017}. To reduce the number of simulated particles, Yano proposed to replace the post-collision VDF by the equilibrium VDF~\cite{Yano2017JCP}. However, in this way, the DSMC solves the Uehling-Uhlenbeck model equation rather than the original QBE, which may introduce large errors when the system is far away from equilibrium as typically occurs in modern experiments~\cite{Vogt2012,WuleiEPL2012,WuleiPRA2012}. It is even surprising that the shear viscosity obtained from the Uehling-Uhlenbeck model equation is smaller than the variational solution that always predicts the lower bound of the transport coefficients.

In recently years, the fast spectral method (FSM), which employs a Fourier-Galerkin discretization in velocity space and handles binary collisions in the corresponding frequency space, has attracted much attention due to its spectral accuracy in solving the Boltzmann collision operator for classical gases~\cite{Mouhot2006}. It  has been successfully applied to calculate the transport coefficients of gas interacting through the Lennard-Jones potentials~\cite{Lei_POF2015}, the  Couette/Poiseuille/thermal transpiration flows~\cite{lei,lei_jfm1,Wu2017JCP_iterative}, linear oscillatory flows in the rectangular cavity~\cite{WUjfm2014,WuPRE2016}, and the spectrum of Rayleigh-Brillouin scattering of the laser-gas interaction~\cite{Wu2014}, and so on. It has also been extended to solve the Boltzmann equation for gas mixtures~\cite{Lei2015FSM,MinhTuanHo2016}, the Enskog equation for dense gases~\cite{Lei2015Enskog,WuJFM2016}, and the  QBE~\cite{Filbet2012a,Hu2012}.

In many recent experiments quantum gas mixtures, either from different species or from different quantum states of the same species, are used~\cite{Somme2011, Sommer2011b,Vogt2012, Koschorreck2013}. However, very few numerical methods are developed for quantum gas mixtures. In this paper we will propose an efficient and accurate FSM to solve the QBE for gas mixtures.

The rest of this paper is organized as follows. In Sec.~\ref{quantumBoltzmannEquation}, the QBE and the equilibrium properties of quantum systems are introduced. In Sec.~\ref{FSMforQuantum}, the FSM to solve the Boltzmann collision operator with general forms of the differential cross-section is presented. Spatially-homogeneous relaxation problems are investigated and factors that affect the accuracy of the FSM are identified in Sec.~\ref{maxwell_relaxation}. In Sec.~\ref{transportCoe}, the accuracy of the FSM is further assessed, by comparing the transport coefficients with the variational solutions. In Sec.~\ref{ConFuture}, we conclude with a summary of the proposed numerical method, and outline future perspectives.

%
%
%

%
%

\section{The quantum Boltzmann equation of gas mixtures}\label{quantumBoltzmannEquation}

Consider a system of quantum gas mixtures in the normal phase, so that it can be described semi-classically by the one-particle VDF  $f^{\imath}(t,\textbf{x},\textbf{v})$, where  $\imath$ denotes the $\imath$-th component, $t$ is the time, $\textbf{x}$ is the spatial coordinate, and $\textbf{v}$ is the molecular velocity. Since the VDF is defined in the way that $(m^\imath/2\pi\hbar)^{d_v}f^{\imath}(t,\textbf{x},\textbf{v})d\textbf{x}d\textbf{v}$ is the the number of the $\imath$-th molecules at time $t$ in the phase-space $d\textbf{x}d\textbf{p}/(2\pi\hbar)^{d_v}= 
(m^\imath/2\pi\hbar)^{d_v}d\textbf{x}d\textbf{v}$, macroscopic quantities such as the number density $n$, bulk velocity $\textbf{V}$, shear stress $P_{ij}$, and heat flux $\textbf{Q}$ of each component can be calculated as the moments of the corresponding VDF:  
\begin{equation}\label{macroscopic_quantities}
\begin{aligned}[b]
n^\imath(\textbf{x},t)=&\left(\frac{m^\imath}{2\pi\hbar}\right)^{d_v}\int f^\imath{d\textbf{v}},\quad 
\textbf{V}^\imath(\textbf{x},t)=\left(\frac{m^\imath}{2\pi\hbar}\right)^{d_v}\frac{1}{n^\imath}
\int{}\textbf{v}f^\imath{}d\textbf{v},\\
P_{ij}^\imath(\textbf{x},t)=&\left(\frac{m^\imath}{2\pi\hbar}\right)^{d_v}m^{\imath}\int{}v_{r,i}v_{r,j}f^\imath{}d\textbf{v}, \quad
\textbf{Q}^\imath(\textbf{x},t)=\left(\frac{m^\imath}{2\pi\hbar}\right)^{d_v}\frac{m^\imath}{2}\int\textbf{v}_{r}|\textbf{v}_r|^2f^\imath{}d\textbf{v},
\end{aligned}
\end{equation}
where $m^\imath$ is the mass of the $\imath$-th component, $\hbar$ is Planck's constant divided by $2\pi$, $d_v=2$ or 3 is the dimension of the problem, $\textbf{v}_r=\textbf{v}-\textbf{V}$ is the peculiar velocity, and indexes $i$ and $j$ are Cartesian components of the spatial variable $\textbf{x}$. Note that $\textbf{p}=m^\imath\textbf{v}$ is the momentum of gas molecules; we here use the velocity $\textbf{v}$ instead of $\textbf{p}$ because it will be easier to develop the FSM that is compatible to our previous works~\cite{lei,lei_jfm1,Lei2015FSM,Lei_POF2015,MinhTuanHo2016,Wu2017JCP_iterative}.

\subsection{Quantum Boltzmann equation}

The QBE is derived from a heuristic argument of the classical Boltzmann equation~\cite{Uehling1933}, where the streaming part remains unchanged as compared to that of the classical molecules, while the collision operator is modified by quantum laws. For fermions, the collision probability is reduced if the final state which the collision leads to has already been occupied, due to Pauli's exclusion principle. For bosons, on the contrary, the collision probability is enhanced. The QBE takes the form of~\cite{Uehling1933} 
\begin{equation}\label{Boltzmann_b}
	\frac{\partial f^\imath}{\partial t}+\textbf{v}\cdot\frac{\partial
	f^\imath}{\partial \textbf{x}}-\frac{1}{m^\imath}\frac{\partial U^\imath}{\partial
	\textbf{x}}\cdot\frac{\partial f^\imath}{\partial
	\textbf{v}}=\sum_{\jmath}\mathcal{Q}^{\imath\jmath}(f^\imath,f^\jmath),
\end{equation}
where $U^\imath(\textbf{x},t)$ are the effective potentials acting on  molecules of the $\imath$-th component, $\mathcal{Q}^{\imath\imath}(f^\imath,f^\imath)$ is the self-collision operator between the $\imath$-th component, and $\mathcal{Q}^{\imath\imath}(f^\imath,f^\jmath)$ with $\imath\neq\jmath$ is the cross-collision operator between molecules of the $\imath$-th and $\jmath$-th components. All the collision operators are local in time and space. For simplicity, $t$ and $\textbf{x}$ are omitted in writing the collision operators in the following general form:
\begin{equation}\label{collision_quantum}
\begin{split}
\mathcal{Q}^{\imath\jmath}(f^\imath,f^\jmath)=\left(\frac{m^\jmath}{2\pi\hbar}\right)^{d_v}
\int_{\mathbb{R}^{d_v}}\int_{\mathbb{S}^{d_v-1}}
&|\textbf{u}|\frac{d\sigma^{\imath\jmath}}{d\Omega}
\bigg\{f^\jmath('\textbf{v}^{\imath\jmath}_{\ast})f^\imath('\textbf{v}^{\imath\jmath})[1+\theta_0f^\jmath(\textbf{v}_{\ast})][1+\theta_0f^\imath(\textbf{v})]\\
&-f^\jmath(\textbf{v}_{\ast})f^\imath(\textbf{v})[1+\theta_0f^\jmath('\textbf{v}^{\imath\jmath}_{\ast})][1+\theta_0f^\imath('\textbf{v}^{\imath\jmath})]\bigg\}d\Omega{}d\textbf{v}_\ast,
\end{split}
\end{equation}
where $\textbf{v}$ and $\textbf{v}_\ast$ are the
pre-collision velocities of molecules of sorts $\imath$ and
$\jmath$, respectively, while $'\textbf{v}^{\imath\jmath}$,
$'\textbf{v}^{\imath\jmath}_\ast$ are the corresponding post-collision
velocities. Conservation of momentum and energy yield the
following relations
\begin{equation}\label{collision_velocity}
\begin{aligned}[b]
'\textbf{v}^{\imath\jmath}=
\textbf{v} 
+\frac{m^\jmath}{m^\imath+m^\jmath}(|\textbf{u}|\Omega-\textbf{u}),\quad
'\textbf{v}^{\imath\jmath}_\ast= \textbf{v}_\ast-\frac{m^\imath}{m^\imath+m^\jmath}(|\textbf{u}|\Omega-\textbf{u}),
\end{aligned}
\end{equation}
where $\textbf{u}=\textbf{v}-\textbf{v}_\ast$ is the relative pre-collision velocity, $\Omega$ is the unit vector in the sphere (or a circle when $d_v=2$) $\mathbb{S}^{d_v-1}$ having the same direction as the relative post-collision velocity, and $\theta$ is the deflection angle between the two relative velocities, i.e. $\cos\theta=\Omega\cdot{\textbf{u}}/|\textbf{u}|$,  $0\le\theta\le\pi$. The differential cross-section is given by $d\sigma^{\imath\jmath}/d\Omega$, which is a function of the relative pre-collision velocity and the deflection angle. Finally,  the Boltzmann equation for molecules obeying the classical statistics is recovered when $\theta_0=0$, while $\theta_0=1$ and $\theta_0=-1$ should be chosen for molecules obeying the quantum Bose-Einstein and Fermi-Dirac statistics, respectively.

In the following numerical simulations by FSM, it is convenient to separate the quantum collision operator~\eqref{collision_quantum} into the following quadratic and cubic collision operators (quartic collision operators cancel out with each other)~\cite{Filbet2012a,Hu2012}:
\begin{equation}
\mathcal{Q}^{\imath\jmath}(f^\imath,f^\jmath)=\mathcal{Q}_c^{\imath\jmath}+\theta_0(\mathcal{Q}_1^{\imath\jmath}+\mathcal{Q}_2^{\imath\jmath}-\mathcal{Q}_3^{\imath\jmath}-\mathcal{Q}_4^{\imath\jmath}),
\end{equation} 
where the classical quadratic collision operator is
\begin{equation}\label{classical}
\mathcal{Q}_c^{\imath\jmath}(f^\imath,f^\jmath)=\left(\frac{m^\jmath}{2\pi\hbar}\right)^{d_v}
\int_{\mathbb{R}^{d_v}}\int_{\mathbb{S}^{d_v-1}}
|\textbf{u}|\frac{d\sigma^{\imath\jmath}}{d\Omega}
[f^\jmath('\textbf{v}^{\imath\jmath}_{\ast})f^\imath('\textbf{v}^{\imath\jmath})  -f^\jmath(\textbf{v}_{\ast})f^\imath(\textbf{v})]d\Omega {}d\textbf{v}_\ast,
\end{equation}
and the cubic collision operators are
\begin{equation}\label{cubic_operators}
\begin{aligned}[b]
\mathcal{Q}_1^{\imath\jmath}=&
\left(\frac{m^\jmath}{2\pi\hbar}\right)^{d_v}\int_{\mathbb{R}^{d_v}}\int_{\mathbb{S}^{d_v-1}}|\textbf{u}|\frac{d\sigma^{\imath\jmath}}{d\Omega}
f^\jmath('\textbf{v}^{\imath\jmath}_{\ast})f^\imath('\textbf{v}^{\imath\jmath})f^\jmath(\textbf{v}_{\ast})d\Omega{}d\textbf{v}_\ast, \\
\mathcal{Q}_2^{\imath\jmath}= &
\left(\frac{m^\jmath}{2\pi\hbar}\right)^{d_v}\int_{\mathbb{R}^{d_v}}\int_{\mathbb{S}^{d_v-1}}|\textbf{u}|\frac{d\sigma^{\imath\jmath}}{d\Omega}
f^\jmath('\textbf{v}^{\imath\jmath}_{\ast})f^\imath('\textbf{v}^{\imath\jmath})f^\imath(\textbf{v})d\Omega{}d\textbf{v}_\ast, \\
\mathcal{Q}_3^{\imath\jmath}=&
\left(\frac{m^\jmath}{2\pi\hbar}\right)^{d_v}\int_{\mathbb{R}^{d_v}}\int_{\mathbb{S}^{d_v-1}}|\textbf{u}|\frac{d\sigma^{\imath\jmath}}{d\Omega}
f^\jmath('\textbf{v}^{\imath\jmath}_{\ast})f^\jmath(\textbf{v}_{\ast})f^\imath(\textbf{v})d\Omega{}d\textbf{v}_\ast, \\
\mathcal{Q}_4^{\imath\jmath}=&
\left(\frac{m^\jmath}{2\pi\hbar}\right)^{d_v}\int_{\mathbb{R}^{d_v}}\int_{\mathbb{S}^{d_v-1}}|\textbf{u}|\frac{d\sigma^{\imath\jmath}}{d\Omega}f^\imath('\textbf{v}^{\imath\jmath})f^\jmath(\textbf{v}_{\ast})f^\imath(\textbf{v})d\Omega{}d\textbf{v}_\ast.
\end{aligned}
\end{equation}

\subsection{Equilibrium properties}

Introducing the entropy density function 
$
s=-\sum_{\imath}\left(\frac{m^\imath}{2\pi\hbar}\right)^{d_v}\int{}[f^\imath\ln{}f^\imath-\theta_0(1+\theta_0f^\imath)\ln(1+\theta_0f^\imath)]d\textbf{v}$
to Eq.~\eqref{Boltzmann_b}, one obtain the
equilibrium VDF
\begin{equation}\label{quantum_equilibrium}
f^\imath_{eq}(t,\textbf{x},\textbf{v})=\left\{
\frac{1}{Z^\imath}\exp\left[\frac{m^\imath(\textbf{v}-\textbf{V})^2}{2k_BT}
\right]-\theta_0\right\}^{-1},
\end{equation}
where $Z^\imath(\textbf{x},t)$ is the local fugacity satisfying
\begin{equation}
Z^\imath(\textbf{x},t)=\exp\left[\frac{\mu^\imath(\textbf{x},t)-U^\imath(\textbf{x},t)}{k_BT}\right],
\end{equation} 
with $\mu^\imath$ and $k_B$ being the chemical potential and Boltzmann constant, respectively.

When the quantum system is in equilibrium, we have
\begin{equation}\label{zeroth}
\begin{aligned}[b]
n^\imath=\left(\frac{m^\imath{}k_BT}{2\pi\hbar^2}\right)^{d_v/2}{G}_{{d_v}/{2}}(Z^\imath), \quad P^\imath_{ij}=n^\imath{k_BT}\frac{{G}_{{d_v}/{2}+1}(Z^\imath)}{{G}_{{d_v}/{2}}(Z^\imath)}\delta_{ij},
\end{aligned}
\end{equation}
where $\delta_{ij}$ is
the Kronecker's delta function, and 
$
{G}_n(Z)=\frac{1}{\Gamma(n)}\int_0^{\infty}\frac{y^{n-1}}{Z^{-1}e^y-\theta_0}dy$
is the Bose-Einstein ($\theta_0=1$) or Fermi-Dirac ($\theta_0=-1$) function, with $\Gamma(n)$ being the Gamma function.

It should be noted that, when the fugacity $Z\rightarrow0$, ${G}_n(Z)\rightarrow Z$, the quantum gas is in the near classical limit, where the equilibrium VDF is very close to the Maxwellian equilibrium VDF for classical gases. Moreover, we have $f^\imath\sim{f^\imath_{eq}}\ll1$, so  the behavior of the quantum gas is similar to the classical one as the quantum correction $\theta_0f^\imath$ can be neglected.

\subsection{Linearized collision operators}

In some cases it is useful to calculate the linearized quantum collision operator, for example, to calculate the transport coefficients such as the shear viscosity, thermal conductivity, and diffusion coefficients. When the system slightly deviates from the equilibrium state~\eqref{quantum_equilibrium}, the one-particle VDF can be expressed as
\begin{equation}\label{perturbation}
f^\imath(t,\textbf{x},\textbf{v})=f^\imath_{eq}(\textbf{x},\textbf{v})+h^\imath(t,\textbf{x},\textbf{v}),
\end{equation}
where $h^\imath$ is the disturbance satisfying $|h^\imath/f_{eq}^\imath|\ll1$.

The quantum Boltzmann collision operator~\eqref{collision_quantum} can be linearized into the following form:
\begin{equation}\label{lin_Boltzmann}
\mathcal{L}^{\imath\jmath}(h^\imath,h^\jmath)=\sum_{\jmath} \left[(\mathcal{L}_{c+}^{\imath\jmath}-\mu_{c}^{\imath\jmath}h^\imath)+\theta_0(\mathcal{L}_1^{\imath\jmath}+\mathcal{L}_2^{\imath\jmath}-\mathcal{L}_3^{\imath\jmath}-\mathcal{L}_4^{\imath\jmath}) \right],
\end{equation} 
where $\mathcal{L}_{c+}^{\imath\jmath}$ and $\mu_{c}^{\imath\jmath}$ are respectively the usual gain part and the equilibrium collision frequency in the classical Boltzmann equation that are defined as~\cite{lei_jfm1,Lei_POF2015,MinhTuanHo2016} 
\begin{equation}\label{equilibrium_frequency}
\begin{aligned}[b]
\mathcal{L}_{c+}^{\imath\jmath}=&\left(\frac{m^\jmath}{2\pi\hbar}\right)^{d_v}
\int\int
|\textbf{u}|\frac{d\sigma^{\imath\jmath}}{d\Omega}
[f^\jmath_{eq}('\textbf{v}^{\imath\jmath}_{\ast})h^\imath('\textbf{v}^{\imath\jmath}) 
+h^\jmath('\textbf{v}^{\imath\jmath}_{\ast})f_{eq}^\imath('\textbf{v}^{\imath\jmath}) -h^\jmath(\textbf{v}_{\ast})f_{eq}^\imath(\textbf{v})]d\Omega {}d\textbf{v}_\ast,\\
\mu_{c}^{\imath\jmath}=&\left(\frac{m^\jmath}{2\pi\hbar}\right)^{d_v}
\int\int
|\textbf{u}|\frac{d\sigma^{\imath\jmath}}{d\Omega}
f_{eq}^\jmath(\textbf{v}_{\ast})d\Omega {}d\textbf{v}_\ast,
\end{aligned}
\end{equation}
while the linearized cubic collision operator $\mathcal{L}_1^{\imath\jmath}$ is obtained by replacing  VDFs in $\mathcal{Q}_1^{\imath\jmath}$ in Eq.~\eqref{cubic_operators} with $h$ and $f_{eq}$ but only keeping the linear term of $h$, in the following manner:
\begin{equation}
\begin{aligned}[b]
\mathcal{L}_1^{\imath\jmath}=\left(\frac{m^\jmath}{2\pi\hbar}\right)^{d_v}\int_{\mathbb{R}^{d_v}}\int_{\mathbb{S}^{d_v-1}}|\textbf{u}|\frac{d\sigma^{\imath\jmath}}{d\Omega}\bigg[&
h^\jmath('\textbf{v}^{\imath\jmath}_{\ast})f_{eq}^\imath('\textbf{v}^{\imath\jmath})f_{eq}^\jmath(\textbf{v}_{\ast}) 
+f_{eq}^\jmath('\textbf{v}^{\imath\jmath}_{\ast})h^\imath('\textbf{v}^{\imath\jmath})f_{eq}^\jmath(\textbf{v}_{\ast})\\
&+f_{eq}^\jmath('\textbf{v}^{\imath\jmath}_{\ast})f_{eq}^\imath('\textbf{v}^{\imath\jmath})h^\jmath(\textbf{v}_{\ast})
\bigg]d\Omega{}d\textbf{v}_\ast,
\end{aligned}
\end{equation}
and the rest cubic collision operators $\mathcal{L}_{2}^{\imath\jmath}$, $\mathcal{L}_{3}^{\imath\jmath}$, and $\mathcal{L}_{4}^{\imath\jmath}$ can be obtained in the same way. It is also obvious that these linearized collision operators can be solved by in the same way as that for the full collision operators.

\section{Fast spectral method for the quantum Boltzmann collision operator}\label{FSMforQuantum}

The approximation of the self-collision quadratic operator~\eqref{classical} (i.e. $\mathcal{Q}_c^{\imath\jmath}$ with $\imath=\jmath$) by the FSM has been studied extensively~\cite{Mouhot2006,Filbet2006,lei,lei_jfm1}, even for generalized forms of the differential cross-section corresponding to general intermolecular potentials such as the Lennard-Jones potential~\cite{Lei_POF2015,WuJFM2017Boundary,Wu2017JCP_iterative}. The approximation of the cubic collision operators~\eqref{cubic_operators} with $\imath=\jmath$ by the FSM has been developed~\cite{Filbet2012a,Hu2012}, while the approximation for the cross-collision operator for classical gas mixtures (i.e. $\mathcal{Q}_c^{\imath\jmath}$ with $\imath\neq\jmath$ and $m^\imath\neq{m^\jmath}$) by the FSM has been recently developed by the present author~\cite{Lei2015FSM,MinhTuanHo2016,Wu2017JCP_iterative}. In this section, on the basis of all these numerical methods, we will develop a FSM for the quantum Boltzmann collision operators with general forms of the differential cross-section, for quantum gas mixtures with different molecular masses. We consider the cross-collision operators~\eqref{collision_quantum} between molecules of the $\imath$-th and $\jmath$-th components only.

\subsection{Carleman-like representation of the collision operator}

As usual, we rewrite the collision operators in Eqs.~\eqref{classical} and~\eqref{cubic_operators} using the Carleman-like representation. With the following basic identity
$
2^{d_v-1}{|\textbf{u}|^{2-d_v}}\int_{\mathbb{R}^{d_v}}\delta(\textbf{y}\cdot{\textbf{u}}+|\textbf{y}|^2)f(\textbf{y})d\textbf{y}
={|\textbf{u}|^{d_v-2}}\int_{\mathbb{S}^{d_v-1}}f\left(\frac{|\textbf{u}|\Omega-\textbf{u}}{2}\right)d\Omega
$,
where $\delta$ is the Dirac delta function, 
the collision operator $\mathcal{Q}_c^{\imath\jmath}$ in Eq.~\eqref{classical} becomes (detailed derivation can be found in Refs.~\cite{lei,Lei2015FSM}): 
\begin{equation}\label{carleman}
\begin{aligned}[b]
\mathcal{Q}_c^{\imath\jmath}
=\int_{\mathbb{R}^{d_v}}\int_{\mathbb{R}^{d_v}}
B^{\imath\jmath}\delta(\textbf{y}\cdot{}\textbf{z})
[f^\jmath(\textbf{v}+\textbf{z}+b^{\imath\jmath}\textbf{y})f^\imath(\textbf{v}+a^{\imath\jmath}\textbf{y})
-f^\jmath(\textbf{v}+\textbf{y}+\textbf{z})f^\imath(\textbf{v})]d\textbf{y}d\textbf{z},
\end{aligned}
\end{equation}
with
\begin{equation}
\begin{aligned}[b]
a^{\imath\jmath}=\frac{2m^\jmath}{m^\imath+m^\jmath}, \quad 
b^{\imath\jmath}=\frac{m^\jmath-m^\imath}{m^\imath+m^\jmath}.
\end{aligned}
\end{equation}

Note that in the derivation of Eq.~\eqref{carleman} we have used the transformations $y=(|\textbf{u}|\Omega-\textbf{u})/2$ and $\textbf{z}=\textbf{v}_\ast-\textbf{v}-\textbf{y}=-\textbf{u}-\textbf{y}$. Therefore, $\textbf{u}=-\textbf{y}-\textbf{z}$ and the deflection angle $\theta$ satisfies $\cos\theta={\Omega\cdot{\textbf{u}}}/{|\textbf{u}|}={-(\textbf{y}-\textbf{z})\cdot(\textbf{y}+\textbf{z})}/{|\textbf{y}+\textbf{z}|^2}$. Note that the delta function $\delta(\textbf{y}\cdot{}\textbf{z})$ poses the condition that the vector $\textbf{z}$ should be perpendicular to the vector $\textbf{y}$, we have $\cos\theta=(|\textbf{z}|^2-|\textbf{y}|^2)/(|\textbf{y}|^2+|\textbf{z}|^2)$ and  $
\theta=2\text{arctan}\left({|\textbf{y}|}/{|\textbf{z}|}\right)$. Since the differential cross-section $d\sigma^{\imath\jmath}/d\Omega$ is a function of the relative pre-collision velocity $|\textbf{u}|$ and the deflection angle $\theta$, $B^{\imath\jmath}$ can be expressed as a function of $|y|$ and $|z|$ only:
\begin{equation}\label{B_yz}
B^{\imath\jmath}=\left(\frac{m^\jmath}{2\pi\hbar}\right)^{d_v}
2^{d_v-1}|\textbf{u}|^{3-d_v}\frac{d\sigma^{\imath\jmath}}{d\Omega}\equiv{B^{\imath\jmath}}(|\textbf{y}|,|\textbf{z}|).
\end{equation}

In numerical calculations, suppose the distribution functions have the support $S$, the relative velocity satisfies $|\textbf{u}|=|\textbf{y}+\textbf{z}|\le2S$, which leads to $|\textbf{y}|,|\textbf{z}|\le{R}=\sqrt{2}S$. Therefore, the infinite integration region with respect to $\textbf{y}$ and $\textbf{z}$ is reduced to
$\mathcal{B}_R$ (i.e. a sphere when $d_v=3$ or a disk when $d_v=2$ of radius $R$ centered on the origin). Consequently, the collision operator in Eq.~\eqref{carleman} is truncated into the following form:
\begin{equation}\label{truncated_quadratic}
\mathcal{Q}_c^{\imath\jmath}=\int_{\mathcal{B}_R}\int_{\mathcal{B}_R}B^{\imath\jmath}(|\textbf{y}|,|\textbf{z}|)\delta(\textbf{y}\cdot\textbf{z})
[f^\jmath(\textbf{v}+\textbf{z}+b^{\imath\jmath}\textbf{y})f^\imath(\textbf{v}+a^{\imath\jmath}\textbf{y})
-f^\jmath(\textbf{v}+\textbf{y}+\textbf{z})f^\imath(\textbf{v})]d\textbf{y}d\textbf{z}.
\end{equation}

Similarly, the cubic collision operators~\eqref{cubic_operators} are transformed and truncated as
\begin{equation}\label{cubic_2}
\begin{aligned}[b]
\mathcal{Q}_1^{\imath\jmath}&=\int_{\mathcal{B}_R}\int_{\mathcal{B}_R}B^{\imath\jmath}(|\textbf{y}|,|\textbf{z}|)\delta(\textbf{y}\cdot\textbf{z}) 
f^\jmath(\textbf{v}+\textbf{z}+b^{\imath\jmath}\textbf{y})f^\imath(\textbf{v}+a^{\imath\jmath}\textbf{y}) f^\jmath(\textbf{v}+\textbf{y}+\textbf{z})d\textbf{y}d\textbf{z}, \\
\mathcal{Q}_2^{\imath\jmath}&=\int_{\mathcal{B}_R}\int_{\mathcal{B}_R}B^{\imath\jmath}(|\textbf{y}|,|\textbf{z}|)\delta(\textbf{y}\cdot\textbf{z})  f^\jmath(\textbf{v}+\textbf{z}+b^{\imath\jmath}\textbf{y})f^\imath(\textbf{v}+a^{\imath\jmath}\textbf{y}) f^\imath(\textbf{v})d\textbf{y}d\textbf{z}, \\
\mathcal{Q}_3^{\imath\jmath}&=\int_{\mathcal{B}_R}\int_{\mathcal{B}_R}B^{\imath\jmath}(|\textbf{y}|,|\textbf{z}|)\delta(\textbf{y}\cdot\textbf{z})  f^\jmath(\textbf{v}+\textbf{z}+b^{\imath\jmath}\textbf{y}) f^\jmath(\textbf{v}+\textbf{y}+\textbf{z})f^\imath(\textbf{v})d\textbf{y}d\textbf{z}, \\
\mathcal{Q}_4^{\imath\jmath}&=\int_{\mathcal{B}_R}\int_{\mathcal{B}_R}B^{\imath\jmath}(|\textbf{y}|,|\textbf{z}|)\delta(\textbf{y}\cdot\textbf{z})  f^\imath(\textbf{v}+a^{\imath\jmath}\textbf{y})   f^\jmath(\textbf{v}+\textbf{y}+\textbf{z})f^\imath(\textbf{v})d\textbf{y}d\textbf{z}.
\end{aligned}
\end{equation}

\subsection{Fast spectral method for truncated collision operators}
	
In FSM, VDFs are periodized on the velocity domain $\mathcal{D}_L=[-L,L)^{d_v}$, where the velocity bound $L$ is chosen to be $L=(3+\sqrt2)S/2$ to avoid the aliasing error caused in the periodization of VDFs and collision operators~\cite{Pareschi2000}. In the Fourier spectral method, VDFs are approximated by the truncated Fourier series,
\begin{equation}\label{inverse_fourier}
\begin{aligned}[b]
f^{\imath}(\textbf{v})=\sum_{j}\hat{f}^{\imath}(\xi_\textbf{j})\exp(i\xi_\textbf{j}\cdot{}\textbf{v}), \quad
\hat{f}^{\imath}(\xi_\textbf{j})=\frac{1}{(2L)^3}
\int_{\mathcal{D}_L}f^{\imath}(\textbf{v})\exp(-i\xi_\textbf{j}\cdot{}\textbf{v})d\textbf{v},
\end{aligned}
\end{equation}
where $i$ is the imaginary unit, and the frequency components are denoted by
\begin{equation}\label{fre_components}
{\xi}=(\xi_1,\xi_2,\cdots,\xi_{d_v})=(j_1,j_2,\cdots,j_{d_v})\frac{\pi}{L}=\textbf{j}\frac{\pi}{L},
\end{equation} 
with $j_k\in[-N_k/2, -N_k/2+1,\cdots, N_k/2-1]$ and $N_k$ being the number of frequency components in the $k$-th direction.

Expanding the truncated collision operators~\eqref{truncated_quadratic} in the truncated Fourier series, we find that the $\textbf{j}$-th model $\widehat{\mathcal{Q}}_c^{\imath\jmath}(\xi_\textbf{j})$ is related to the Fourier coefficients $\hat{f}^\imath$ and $\hat{f}^\jmath$ as
\begin{equation}\label{mode_binary}
\widehat{\mathcal{Q}}_c^{\imath\jmath}(\xi_\textbf{j})= \sum_{\textbf{l}+\textbf{m}=\textbf{j} \atop
	\textbf{l},\textbf{m}}
\hat{f}^\imath_\textbf{l}\hat{f}^\jmath_\textbf{m}\beta(a\xi_\textbf{l}+b\xi_\textbf{m},\xi_\textbf{m})
-\hat{f}^\imath_\textbf{l}\hat{f}^\jmath_\textbf{m}\beta(\xi_\textbf{m},\xi_\textbf{m}).
\end{equation}
Similarly, the $\textbf{j}$-th mode of the truncated cubic collision operators~\eqref{cubic_2} are expressed as
\begin{equation}\label{cubic_carlemann}
\begin{aligned}[b]
\widehat{\mathcal{Q}}_1^{\imath\jmath}(\xi_\textbf{j})=& 		 \sum_{\textbf{l}+\textbf{m}+\textbf{n}=\textbf{j} \atop   \textbf{l},\textbf{m},\textbf{n}}
\hat{f}^\imath_\textbf{l}\hat{f}^\jmath_\textbf{m}\hat{f}^\jmath_\textbf{n}
\beta(a\xi_\textbf{l}+b\xi_\textbf{m}+\xi_\textbf{n},\xi_\textbf{m}+\xi_\textbf{n}), \ \
\widehat{\mathcal{Q}}_2^{\imath\jmath}(\xi_\textbf{j})= \sum_{\textbf{l}+\textbf{m}+\textbf{n}=\textbf{j} \atop   \textbf{l},\textbf{m},\textbf{n}}\hat{f}^\imath_\textbf{l}\hat{f}^\jmath_\textbf{m}\hat{f}^\imath_\textbf{n}
\beta(a\xi_\textbf{l}+b\xi_\textbf{m},\xi_\textbf{m}), \\
\widehat{\mathcal{Q}}_3^{\imath\jmath}(\xi_\textbf{j})=& \sum_{\textbf{l}+\textbf{m}+\textbf{n}=\textbf{j} \atop   \textbf{l},\textbf{m},\textbf{n}}\hat{f}^\imath_\textbf{l}\hat{f}^\jmath_\textbf{m}\hat{f}^\imath_\textbf{n}
\beta(\xi_\textbf{m}+a\xi_\textbf{n},\xi_\textbf{m}), \ \widehat{\mathcal{Q}}_4^{\imath\jmath}(\xi_\textbf{j})= \sum_{\textbf{l}+\textbf{m}+\textbf{n}=\textbf{j} \atop
	\textbf{l},\textbf{m},\textbf{n}}\hat{f}^\imath_\textbf{l}\hat{f}^\jmath_\textbf{m}\hat{f}^\jmath_\textbf{n}
\beta(\xi_\textbf{m}+b\xi_\textbf{n},\xi_\textbf{m}+\xi_\textbf{n}),
\end{aligned}
\end{equation}
where the kernel mode $\beta(\textbf{l},\textbf{m})$ is
\begin{equation}\label{kernel_mode0}
\beta(\xi_\textbf{l},\xi_\textbf{m})=\int_{\mathcal{B}_R}\int_{\mathcal{B}_R}{}B^{\imath\jmath}(|\textbf{x}|,|\textbf{y}|)
\delta(\textbf{y}\cdot{\textbf{z}})
\exp(i\xi_\textbf{l}\cdot{\textbf{y}}+i\xi_\textbf{m}\cdot{\textbf{z}})d\textbf{y}d\textbf{z}.
\end{equation}

Note that the second term on the right-hand side of Eq.~\eqref{mode_binary} can be calculated by the FFT-based convolution with the computational cost $O(N^{d_v}\log{N})$, where $N$ is at the same order of the number of frequency components $N_k$ in Eq.~\eqref{fre_components}. For the first term on the right-hand size of Eq.~\eqref{mode_binary}, however, the direct calculation requires a computational cost of $O(N^{2d_v})$. Direct calculations of the cubic collision operators in Eq.~\eqref{cubic_carlemann} are even more time-consuming, at the order of $N^{3d_v}$. Our main goal is to separate $\xi_\textbf{l}$ and $\xi_\textbf{m}$ in the kernel mode $\beta(\xi_\textbf{l},\xi_\textbf{m})$ so that Eqs.~\eqref{mode_binary} and~\eqref{cubic_carlemann} can be calculated effectively by the FFT-based convolution, with a relative low computational cost.

\subsubsection{Approximation of the kernel mode}
	
Introducing the transforms $\textbf{y}=\rho\textbf{e}$ and $\textbf{z}=\rho'\textbf{e}'$, 	where $\textbf{e}$ and $\textbf{e}'$ are vectors in the unit sphere when $d_v=3$ and unit circle when $d_v=2$, the kernel mode~\eqref{kernel_mode0} is expressed in the spherical ($d_v=3$) or polar ($d_v=2$) coordinates as
\begin{equation}\label{kernel_model_GL}
\begin{aligned}[b]
&\int\int \delta(\textbf{e}\cdot\textbf{e}')
\int_{0}^R\int_{0}^R(\rho\rho')^{d_v-2}B^{\imath\jmath}(\rho,\rho')
\exp(i\rho\xi_\textbf{l}\cdot{\textbf{e}})
\exp(i\rho'\xi_\textbf{m}\cdot{}\textbf{e}')
{d\rho'}d\rho{}d\textbf{e}'d\textbf{e}\\
=&\sum_{r=1}^{M_2}\int\int \delta(\textbf{e}\cdot\textbf{e}')
\int_{0}^R\omega_r(\rho_r\rho')^{d_v-2}B^{\imath\jmath}(\rho_r,\rho')
\exp(i\rho_r\xi_\textbf{l}\cdot{\textbf{e}})
\exp(i\rho'\xi_\textbf{m}\cdot{}\textbf{e}')
{d\rho'}{}d\textbf{e}'d\textbf{e}\\
=&\sum_{r=1}^{M_2}\int\int \delta(\textbf{e}\cdot\textbf{e}')
\exp(i\rho_r\xi_\textbf{l}\cdot{\textbf{e}})
\phi(\rho'_r,\xi_\textbf{m}\cdot{}\textbf{e}')
d\textbf{e}'d\textbf{e},
\end{aligned}
\end{equation}
where the integral with respect to $\rho$ has been approximated by Gauss-Legendre quadrature, with $\rho_r$ and $\omega_r$ ($r=1,2,\cdots,M_2$) being respectively the abscissas and weights of the Gauss-Legendre quadrature in the region of $0\le\rho\le{}R$, and the term
\begin{equation}\label{phiphi}
\phi(\rho'_r,\xi_\textbf{m}\cdot{}\textbf{e}')=\int_{0}^R \omega_r(\rho_r\rho')^{d_v-2}B^{\imath\jmath}(\rho_r,\rho')\cos(\rho'\xi_\textbf{m}\cdot{}\textbf{e}'){d\rho'},
\end{equation}
can be calculated accurately by some high order numerical quadrature. 

It should be noted that the term $\rho_r\xi_\textbf{l}\cdot{\textbf{e}}$ in Eq.~\eqref{kernel_model_GL} is at the order of $N$ for the largest $\rho_r=R$ and the largest frequency $\xi_\textbf{l}=\pi/L$. Therefore, the function $\exp(i\rho_r\xi_\textbf{l}\cdot{\textbf{e}})$ oscillates maximally $O(N)$ times. Consequently, $M_2$ should be roughly of the order of $N$ to make the integral with respect to $\rho$ in Eq.~\eqref{kernel_model_GL} by Gauss-Legendre quadrature accurate. In practical calculation, however, since the spectra of the VDF at high frequency components are very small, $M_2$ can be several times smaller than $N$ to have better numerical efficiency; this point will be demonstrated in the numerical simulation in Sec.~\ref{maxwell_relaxation}. Also, note that in the evaluation of the integral with respect to $\rho'$, the imaginary part is omitted due to the symmetry condition, that is, $B^{\imath\jmath}$, which is related to the differential cross-section, remains unchanged when $\textbf{e}'$ is replaced by $-\textbf{e}'$, see Eq~\eqref{B_yz}.

After some algebraic manipulation (see descriptions from Eq.~(34) to Eq.~(38) in Ref.~\cite{lei} when $d_v=3$, and Eqs.~(15) and (16) in Ref.~\cite{Lei2015Enskog} when $d_v=2$), we have
\begin{itemize}
	\item 
	when $d_v=3$, the integral with respect to the unit vector $\textbf{e}$ in a sphere is approximated by the trapezoidal rule, i.e. $e_{\theta_p,\varphi_q}=(\sin\theta_p\cos\varphi_q,\sin\theta_p\sin\varphi_q,\cos\theta_p)$ with 					$\theta_p=p\pi/M$ and $\varphi_q=q\pi/M$, where $p,q=1,2,\cdots, M$), and the kernel mode~\eqref{kernel_model_GL} can be approximated by:
	\begin{equation}\label{kernelmode3d}
	\beta(\textbf{l},\textbf{m})
	\simeq\frac{2\pi^2}{M^2}\sum_{r,p,q=1}^{M_2,M-1,M}
	\cos(\rho_r\xi_\textbf{l}\cdot{\textbf{e}_{\theta_p,\varphi_q}})
	\psi_3\left(\rho_r,\sqrt{|\xi_\textbf{m}|^2-(\xi_\textbf{l}\cdot{\textbf{e}_{\theta_p,\varphi_q}})^2}\right)\sin\theta_p,
	\end{equation}
	where  
	$
	\psi_3(\rho_r,s)= 2\pi\int_0^R \omega_r{\rho_r\rho'}B^{\imath\jmath}(\rho_r,\rho')
	J_0(\rho' s)d\rho'$,
	with $J_0$ being the zeroth-order Bessel function of  first kind.
	
	\item 
	when $d_v=2$, the integral with respect to the unit vector $\textbf{e}$ in a circle is approximated by the trapezoidal rule, i.e. $e_{\theta_p}=(\cos\theta_p,\sin\theta_p)$ with
	$\theta_p=p\pi/M$, where $p=1,2,\cdots, M$), and the kernel mode~\eqref{kernel_model_GL} is approximated by
	\begin{equation}\label{kernelmode2d}
	\begin{aligned}[b]
	\beta(\textbf{l},\textbf{m})&\simeq\frac{\pi}{M}\sum_{r,p=1}^{M_2,M} \cos(\rho_r\xi_\textbf{l}\cdot{\textbf{e}_{\theta_p}})
	\psi_2(\rho_r,\xi_\textbf{m}\cdot{\textbf{e}_{\theta_p+\frac{\pi}{2}}}),
		\end{aligned}
	\end{equation}
	where
	$
	\psi_2(\rho_r,s)=4\int_{0}^R \omega_rB^{\imath\jmath}(\rho_r,\rho')\cos(\rho'\xi_\textbf{m}\cdot{}\textbf{e}'){d\rho'}$.
	
	
\end{itemize}

From Eqs.~\eqref{kernelmode3d} and~\eqref{kernelmode2d}, we see that $\xi_\textbf{l}$ and $\xi_\textbf{m}$ are separated in two different functions, which enables faster computation of the quantum collision operator via the FFT-based convolution. The major algorithm is described below.

\subsection{Detailed numerical implementation}\label{detailed_imp}		

We take the 2D case as an example to demonstrate how the FSM is implemented. First the cosine function in Eq.~\eqref{kernelmode2d} is expressed in terms of the exponential function:
\begin{equation}
\cos(\rho_r\xi_\textbf{l}\cdot{\textbf{e}_{\theta_p}})=\frac{\exp(i\rho_r\xi_\textbf{l}\cdot{\textbf{e}_{\theta_p}})
+\exp(-i\rho_r\xi_\textbf{l}\cdot{\textbf{e}_{\theta_p}})}{2},
\end{equation}
and for simplicity only the term $\exp(i\rho_r\xi_\textbf{l}\cdot{\textbf{e}_{\theta_p}})$ is considered in the following paper, as the term $\exp(-i\rho_r\xi_\textbf{l}\cdot{\textbf{e}_{\theta_p}})$ can be handled similarly.

The spectrum of the quadratic collision operator~\eqref{mode_binary}, can be expressed as
\begin{equation}\label{Qc_detailed}
\begin{aligned}[b]
\widehat{\mathcal{Q}}_c^{\imath\jmath}(\xi_\textbf{j})\simeq &\frac{\pi}{M}\sum_{r,p=1}^{M_2,M}\sum_{\textbf{l}+\textbf{m}=\textbf{j} \atop    \textbf{l},\textbf{m}}
\underbrace{  \exp(ia\rho_r\xi_\textbf{l}\cdot{\textbf{e}_{\theta_p}})\hat{f}^\imath_\textbf{l}
	\times
	\exp(ib\rho_r\xi_\textbf{m}\cdot{\textbf{e}_{\theta_p}})
	\psi_2(\rho_r,\xi_\textbf{m}\cdot{\textbf{e}_{\theta_p+\frac{\pi}{2}}})
\hat{f}^\jmath_\textbf{m} }_{C_1^{rp}(l+m)}\\
-&\frac{\pi}{M}\sum_{\textbf{l}+\textbf{m}=\textbf{j} \atop
	\textbf{l},\textbf{m}}
\hat{f}^\imath_\textbf{l} \times\sum_{r,p=1}^{M_2,M}
{\exp(i\rho_r\xi_\textbf{m}\cdot{\textbf{e}_{\theta_p}})
	\psi_2(\rho_r,\xi_\textbf{m}\cdot{\textbf{e}_{\theta_p+\frac{\pi}{2}}})
\hat{f}^\jmath_\textbf{m}},
\end{aligned}
\end{equation}
where the term $C_1^{rp}(l+m)$ is a convolution between the function $\exp(ia\rho_r\xi_\textbf{l}\cdot{\textbf{e}_{\theta_p}})\hat{f}^\imath_\textbf{l}$ and $\exp(ib\rho_r\xi_\textbf{m}\cdot{\textbf{e}_{\theta_p}})
\psi_2(\rho_r,\xi_\textbf{m}\cdot{\textbf{e}_{\theta_p+\frac{\pi}{2}}})
\hat{f}^\jmath_\textbf{m}$. This term needs to be evaluated $MM_2$ times, each with a computational cost of $O(N^2\log{N})$. Therefore, the total computational cost for the term $\widehat{\mathcal{Q}}_c^{\imath\jmath}(\xi_\textbf{j})$ should be $O(MM_2N^2\log{N})$, which is at the order of $MN^3\log{N}$ since $M_2\sim{N}$, see the paragraph after Eq.~\eqref{phiphi}.

When $C_1^{rp}$ in Eq.~\eqref{Qc_detailed} is obtained, the spectrum of the cubic collision operator $\mathcal{Q}_2^{\imath\jmath}$ can be expressed as
\begin{equation}
\widehat{\mathcal{Q}}_2^{\imath\jmath}(\xi_\textbf{j})\simeq
\frac{\pi}{M}\sum_{n}
\hat{f}^\imath_\textbf{n}\times{} \sum_{r,p=1}^{M_2,M}C_1^{rp}(\textbf{j}-\textbf{n}), 
\end{equation}
which can be solved by the FFT-based convolution with a computational cost of $O(N^2\log{N})$, that is, the cost is negligible when compared to Eq.~\eqref{Qc_detailed}.

The spectrum of the cubic collision operators $\mathcal{Q}_1$ can be expressed as
\begin{equation}\label{c2rp}
\begin{aligned}[b]
\widehat{\mathcal{Q}}_1^{\imath\jmath}(\xi_\textbf{j})\simeq&
\frac{\pi}{M}\sum_{r,p=1}^{M_2,M}\sum_{l}
{\exp(ia\rho_r\xi_l\cdot{\textbf{e}_{\theta_p}})\hat{f}^\imath_\textbf{l}}\\
&\times\sum_{\textbf{m}+\textbf{n}=\textbf{j}-\textbf{l} \atop   m,n}\underbrace{{{
\exp(ib\rho_r\xi_\textbf{m}\cdot{\textbf{e}_{\theta_p}})\hat{f}^\jmath_\textbf{m}}\times
	{\exp(i\rho_r\xi_\textbf{n}\cdot{\textbf{e}_{\theta_p}})\hat{f}^\jmath_\textbf{n}}}
	\psi_2(\rho_r,\xi_{\textbf{m}+\textbf{n}}\cdot{\textbf{e}_{\theta_p+\frac{\pi}{2}}})}_{C_2^{rp}(\textbf{m}+\textbf{n})},
\end{aligned}
\end{equation}
where the underlined term is a convolution that can be computed via FFT with a cost of $O(N^2\log{N})$, and the result of which multiplied by $\psi_2(\rho_r,\xi_{\textbf{m}+\textbf{n}}\cdot{\textbf{e}_{\theta_p+\frac{\pi}{2}}})$ forms $C_2^{rp}(\textbf{m}+\textbf{n})$. The terms $C_2^{rp}(\textbf{m}+\textbf{n})$ and $\exp(ia\rho_r\xi_l\cdot{\textbf{e}_{\theta_p}})\hat{f}^\imath_\textbf{l}$ form the convolution again, which can be calculated with a cost of  $O(N^2\log{N})$. Since this convolution has to be repeated $MM_2$ times, the total computational cost will be $O(MN^3\log{N})$.

When $C_2^{rp}$ in Eq.~\eqref{c2rp} is obtained, the spectrum of the cubic collision operator $Q_4$ can be expressed as: 
\begin{equation}
\widehat{\mathcal{Q}}_4^{\imath\jmath}(\xi_\textbf{j})\simeq
\frac{\pi}{M}\sum_{l}
\hat{f}^\imath_\textbf{l}\times{}\sum_{r,p=1}^{M_2,M}C_2^{rp}(\textbf{j}-\textbf{l}).
\end{equation}
which can be calculated with the cost $O(N^2\log{N})$.

The spectral of the cubic collision operator $Q_3$, as given in Eq.~\eqref{cubic_carlemann}, can be expressed as:
\begin{equation}
\begin{aligned}[b]
\widehat{\mathcal{Q}}_3^{\imath\jmath}(\xi_\textbf{j})\simeq&
\frac{\pi}{M}\sum_{l}
\hat{f}^\imath_\textbf{l} \sum_{r,p=1}^{M_2,M}\sum_{\textbf{m}+\textbf{n}=\textbf{j}-\textbf{l} \atop
	m,n}
{\exp(ia\rho_r\xi_\textbf{n}\cdot{\textbf{e}_{\theta_p}})\hat{f}^\imath_\textbf{n}}\times{}\exp(i\rho_r\xi_\textbf{m}\cdot{\textbf{e}_{\theta_p}})\hat{f}^\jmath_\textbf{m}\psi_2(\rho_r,\xi_\textbf{m}\cdot{\textbf{e}_{\theta_p+\frac{\pi}{2}}}),
\end{aligned}
\end{equation}
where the computational cost will be $O(MN^3\log{N})$, like $\mathcal{Q}_c^{\imath\jmath}$ and $\mathcal{Q}_1^{\imath\jmath}$.

When $\widehat{Q}^{\imath\jmath}$ is obtained, the collision operator ${Q}^{\imath\jmath}$ can be obtained through following FFT, with a cost $O(N^2\log{N})$:
\begin{equation}
{Q}^{\imath\jmath}(\textbf{v})=\sum_{\textbf{j}}\widehat{Q}^{\imath\jmath}(\xi_\textbf{j})\exp(i\xi_\textbf{j}\cdot\textbf{v}).
\end{equation}

Therefore, if the FFT-based convolution is applied, for the case of $d_v=2$, the overall computational cost is $O(MN^3\log{N})$, while for $d_v=3$, the computational cost is $O(M^2N^4\log{N})$. Note that the procedure in deriving the FSM for quantum Boltzmann equation is essentially the same as that for the classical Boltzmann equation, therefore, it can be proved that the present FSM conserves the mass and satisfies the H-theorem, while errors on the approximations of momentum and energy are spectrally small~\cite{Mouhot2006}.

\section{The spatially-homogeneous relaxation of quantum gases}\label{maxwell_relaxation}

In this section, we assess the performance of FSM in the study of spatially-homogeneous relaxation of binary  gas mixtures of components $A$ and $B$. Since the property of self-collision operators has been well investigated~\cite{lei,Filbet2006,Hu2012},  we focus on the cross-collision collision operators only. This situation actually occurs in Fermi gases where interactions between fermions with the same spin (i.e. described by the self-collision operator) are much smaller than those between opposite spins (i.e. described by the cross-collision operator)~\cite{Vogt2012,bruun_2012,arxiv_sch}. For simplicity, we consider the case of $d_v=2$, with the following differential cross-section~\cite{bruun_2012}:
\begin{equation}\label{two_d_s}
\frac{d\sigma^{\imath\jmath}}{d\Omega}=\frac{2\pi\hbar}{m_r|\textbf{u}|}\frac{1}{\log^2(a^2_sm^2_r|\textbf{u}|^2/\hbar^2)+\pi^2}, 
\end{equation}
where $a_s$ is the s-wave scattering length, which can be controlled experimentally via Feshbach resonance, and $m_r={m^Am^B}/{(m^A+m^B)}$
is the reduced mass.

The evolution of VDFs for components A and B in the spatially-homogeneous relaxation is governed by the following equations  
\begin{equation}\label{Boltzmann_homo}
\begin{aligned}[b]
\frac{\partial f^A}{\partial t'}=Q^{AB}(f^A,f^B),\quad
\frac{\partial f^B}{\partial t'}=Q^{BA}(f^B,f^A),
\end{aligned}
\end{equation}
with the following cross-collision operator
\begin{equation}\label{rescale}
\begin{aligned}[b]
\mathcal{Q}^{\imath\jmath}(f^\imath,f^\jmath)=\left(\frac{m^\jmath}{m^A}\right)^{2}
\int\int
&\frac{d\Omega
d\textbf{v}_\ast}{\log^2(a|\textbf{u}|^2)+\pi^2}
\bigg\{f^\jmath('\textbf{v}^{\imath\jmath}_{\ast})f^\imath('\textbf{v}^{\imath\jmath})[1+\theta_0f^\jmath(\textbf{v}_{\ast})][1+\theta_0f^\imath(\textbf{v})]\\
&-f^\jmath(\textbf{v}_{\ast})f^\imath(\textbf{v})[1+\theta_0f^\jmath('\textbf{v}^{\imath\jmath}_{\ast})][1+\theta_0f^\imath('\textbf{v}^{\imath\jmath})]\bigg\},
\end{aligned}
\end{equation}
where $t'=tm^Ak_BT_r/\pi\hbar{m_r}$, $a=2k_BT_ra_s^2m_r^2/m^A$, and the velocity have been normalized by $\sqrt{2k_BT_{r}/m^A}$, with $T_{r}$ being the reference temperature.  We will study how the initial non-equilibrium VDFs
\begin{equation}\label{initial}
f^A(t=0,\textbf{v})=f^B(t=0,\textbf{v})=\frac{8}{\pi}|\textbf{v}|^2\exp(-|\textbf{v}|^2),
\end{equation} 
relax to the final equilibrium states.

\begin{figure}[t]
	\centering
	\includegraphics[scale=0.55,viewport=70 270 900 500,clip=true]{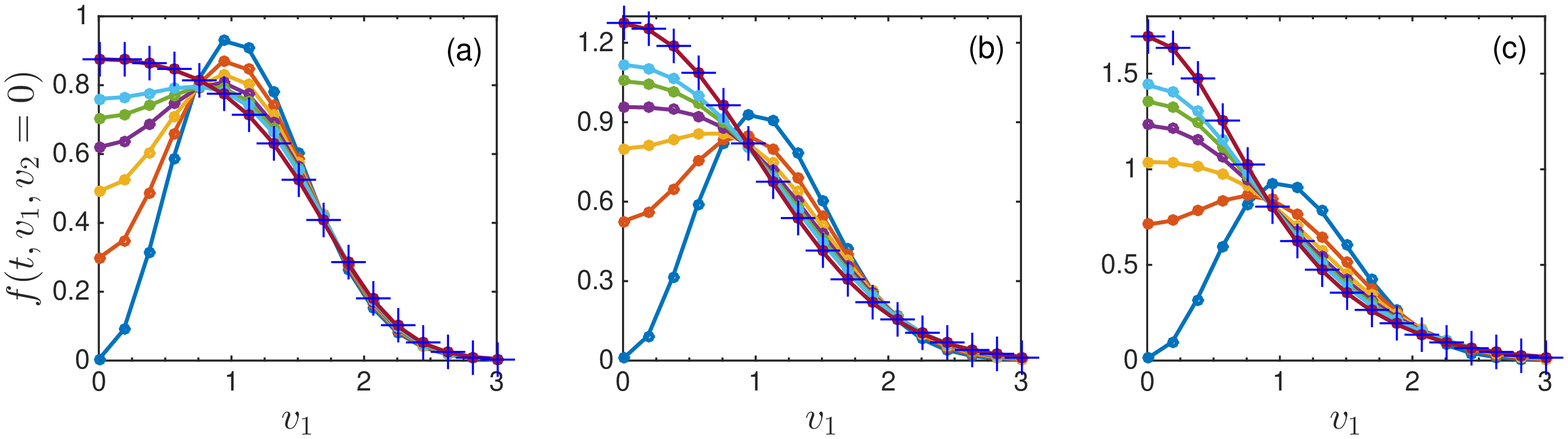}\\
	\vskip 0.5cm
	\includegraphics[scale=0.6,viewport=30 240 620 450,clip=true]{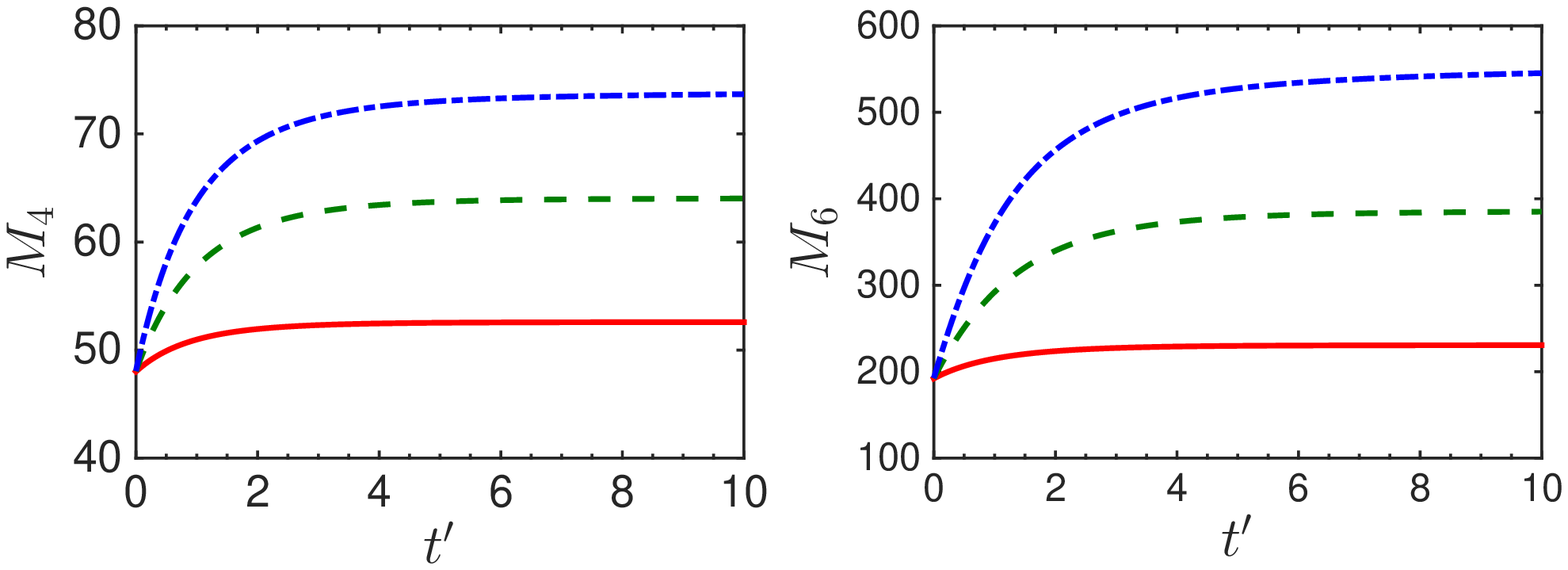}
	\caption{ (top row) The spatially-homogeneous relaxation of VDFs  for (a) Fermi, (b) classical, and (c) Bose gases, where the differential cross-section is given by Eq.~\eqref{two_d_s} with the normalized parameter $a=1$ in Eq.~\eqref{rescale}. Due to symmetry only the region $v_1>0$ is shown. In each figure, from bottom to top (near $v_1=0$), the time $t'$ for each line is 0, 0.25, 0.50, 0.75, 1, 1.25, and 10, respectively. The symbol `crosses' show the equilibrium VDF~\eqref{equi_equal_mass}.  (bottom) The time evolution of the fourth- and sixth-order moments of the VDF: $M_4(t)=\int\int f(\textbf{v},t)|\textbf{v}|^4d\textbf{v}$ and $M_6(t)=\int\int f(\textbf{v},t)|\textbf{v}|^6d\textbf{v}$. Solid, dashed, and dash-dotted lines are the results for Fermi,  classical, and Bose gases. }
	\label{Relax_equalMass}
\end{figure}

\subsection{The equal-mass mixture}\label{equal_mass}

Since the mass and energy are conserved during the collision, for the equal-mass case (i.e. $m^A=m^B$), the final equilibrium state corresponding to the initial condition~\eqref{initial} is
\begin{equation}\label{equi_equal_mass}
f^A(t=\infty,\textbf{v})=f^B(t=\infty,\textbf{v})=\left\{
\frac{1}{Z}\exp\left(\frac{|\textbf{v}|^2}{T}
\right)-\theta_0\right\}^{-1},
\end{equation}
where the equilibrium fugacity and temperature $(Z,T)$ are (7.0363, 1.2219), (1.2732, 2.0000), and (0.6291, 2.5671) for the Fermi, classical, and Bose gases, respectively.

Figure~\ref{Relax_equalMass} depicts the relaxation-to-equilibrium process of VDFs, as well as the time evolution of the fourth- and sixth-order moments, when Eq.~\eqref{Boltzmann_homo} is solved by Euler method with a time-step of $0.0025$, and the collision operator~\eqref{rescale} is approximated by the FSM with the following parameters: the number of solid angle is discretized  uniformly with $M=10$, the velocity domain $[-L,L)^2$ with $L=6$ is discretized by $N=64$ uniform grid points in each direction, and $M_2=64$ is chosen in the Gauss-Legendre approximation used in Eq.~\eqref{kernel_model_GL}. It can be seen from Fig.~\ref{Relax_equalMass}(a,b,c) that the final equilibrium states agree well with the analytical solution~\eqref{equi_equal_mass}. Mathematically it has been proven for classical Boltzmann equation that the FSM preserves the mass accurately while the energy is conserved with spectral accuracy~\cite{Mouhot2006}; from the numerical simulation with the above detailed parameters, these conclusions hold also for the quantum Boltzmann equation as, for example, for Fermi gas the maximum relative variation in mass and energy during the whole relaxation process are $2.7\times10^{-15}$ and $4.4\times10^{-7}$, respectively. Thus, the VDF, as well as its fourth- and sixth-order moments, are chosen as reference solutions to investigate factors that affect the accuracy of the FSM.

\begin{figure}[p]
	\centering	\includegraphics[scale=0.5,viewport=20 0 690 380,clip=true]{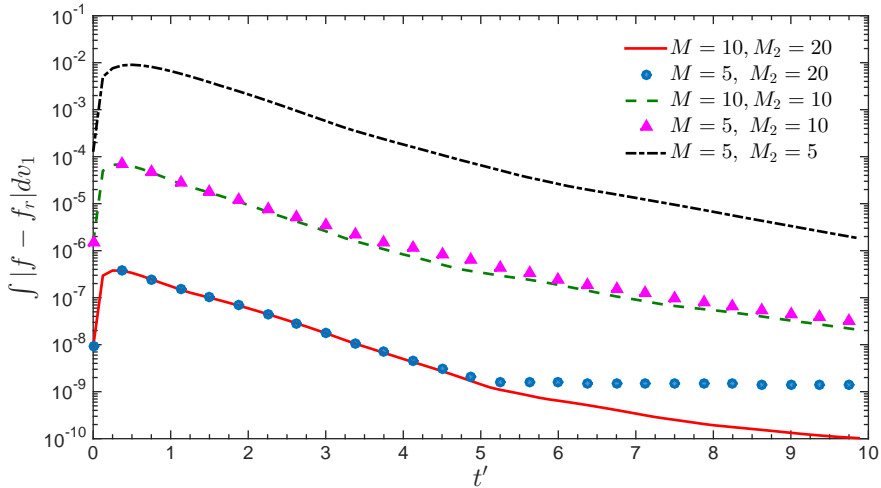}
	\caption{ The relative error in the mesoscopic VDF of Fermi gas evaluated at $v_2=0$, when the velocity space $[-6,6]^2$ is discretized by $64\times64$ uniform points. Note that the reference solution $f_r$ is obtained by FSM with the same parameters as used in Fig.~\ref{Relax_equalMass}. }
	\label{vdf_error}
\end{figure}

\begin{figure}
	\centering
	\includegraphics[scale=0.55,viewport=30 0 850 450,clip=true]{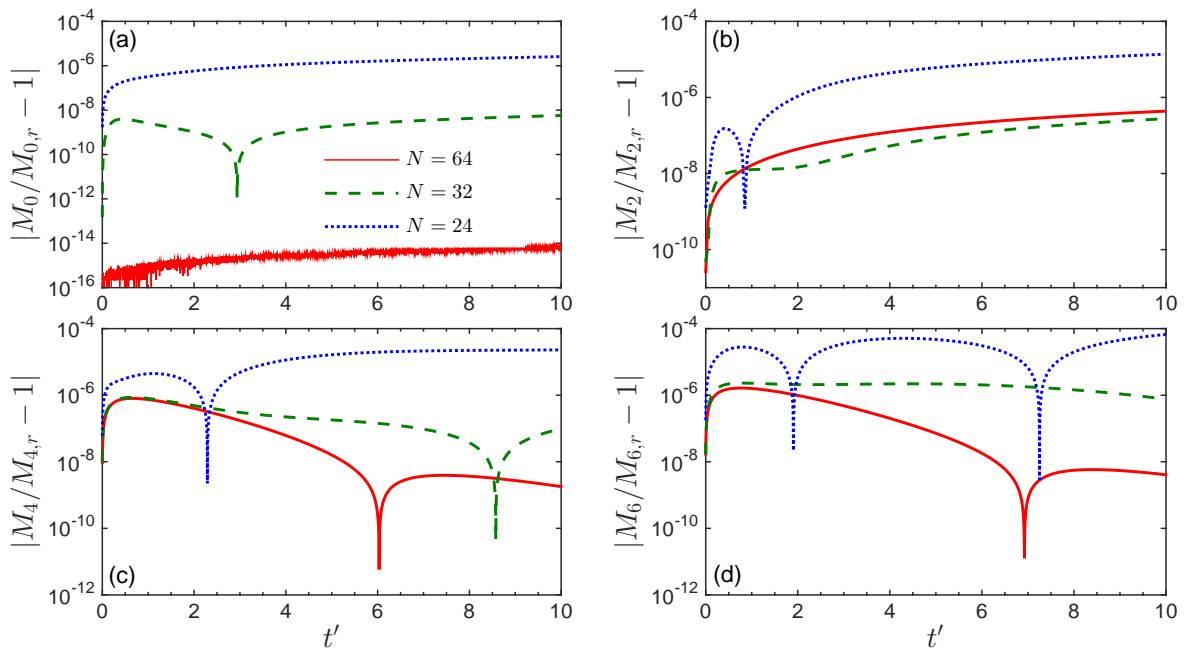}
	\caption{The relative errors of the zeroth-, second-, fourth- and sixth-order moments of the VDF of Fermi gas as compared to the reference solutions with $N=M_2=64$.  $M=5$ and $M_2=10$ are chosen, while other parameters are the same as in Fig.~\ref{Relax_equalMass}. Note that $M_{0,r}$ and $M_{2,r}$ are calculated based on the initial VDF, since the mass and energy is conserved during the homogeneous relaxation.  }
	\label{Relax_equalMass_M4M6_error}
\end{figure}

Figure~\ref{vdf_error} shows the absolute error in the VDF when the velocity grids are kept at $64\times64$, while values of $M$ and $M_2$ are reduced. When the value of $M_2$ is fixed, it is seen that decreasing the number of discrete solid angle $M$ from 10 to 5 only slightly affects the accuracy. Therefore, $M=5$ can be considered accurate, as has been chosen in our previous numerical simulations of the classical Boltzmann equation~\cite{lei,lei_jfm1}. The value of $M_2$, however, strongly affects the accuracy. Theoretically, $M_2$ should be at the order of $N$ to make the approximation in Eq.~\eqref{kernel_model_GL} sufficiently accurate for each frequency component, see the paragraph after Eq.~\eqref{phiphi}. However, at large frequency components the kernel mode $\beta({\textbf{l},\textbf{m}})$ in Eq.~\eqref{kernelmode2d} and the spectrum of the VDF are sufficiently small, therefore, $M_2$ can be smaller than $N$: in Fig.~\ref{vdf_error} it is seen that even $M_2=10$ has good accuracy.

Figure~\ref{Relax_equalMass_M4M6_error} shows the relative errors of the zeroth-, second-, fourth-, and sixth-order moments of the VDF as a function of the time. Odd-order moments are not included because they are zero due to symmetry. From this figure we can see that the accuracy deteriorates when the number of velocity points and frequency components $N^2$ decrease. When $N=64$, from Fig.~\ref{Relax_equalMass_M4M6_error}(a) we find that the mass is conserved to the machine accuracy. However, as $N$ decreases,  the mass is not strictly conserved. For example, when $N=24$. This is because the discretized frequency components does not cover the whole spectrum of the VDF, such that some information is lost, and consequently the mass is not conserved; if higher accuracy is required when $N=24$, the velocity domain should be reduced by decreasing the value of $L$ such that the discretized frequency components will cover the whole spectrum of the VDF, as from Eq.~\eqref{fre_components} we find that the range of the frequency is  inversely proportional to $L$. From Fig.~\ref{Relax_equalMass_M4M6_error}(b)  we see  that the energy (temperature) is not conserved, but the maximum relative deviation from the initial value is about $10^{-5}$ when $N=24$ and $10^{-6}$ when $N=32$. Although the relative error generally increases with the order of moment, deviations of the sixth-order moment from reference solutions are still very small for the parameters considered.

\begin{figure}[t]
	\centering
	\includegraphics[scale=0.46,viewport=90 0 1090 530,clip=true]{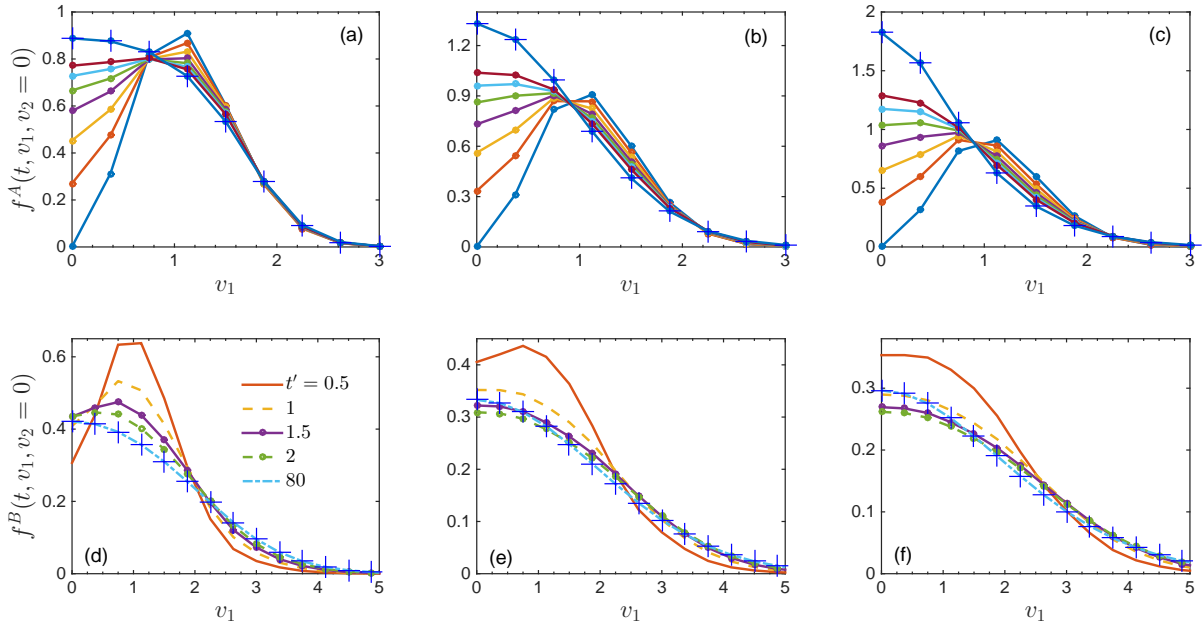}
	\caption{ The relaxation of VDFs in the binary mixture with $m^A=4m^B$. (top row) The spatially-homogeneous relaxation of VDFs $f^A$ for (a) Fermi, (b) classical, and (c) Bose gases, where the differential cross-section is given by Eq.~\eqref{two_d_s} with the normalized parameter $a=1$ in Eq.~\eqref{rescale}. In each figure, from bottom to top (near $v_1=0$), the time $t'$ for each line is 0, 2, 4, 6, 1, 8, 10, 12, and 80, respectively. (bottom row) The spatially-homogeneous relaxation of VDFs $f^B$ for (d) Fermi, (e) classical, and (f) Bose gases. Note that in all the figures, the symbol `crosses' shows the equilibrium VDF given by Eq.~\eqref{unequal_eq}. Due to symmetry only the region $v_1>0$ is shown. }
	\label{Relax_UnequalMass}
\end{figure}

\subsection{The unequal-mass mixture}

We now consider the case of unequal-mass mixture, where the molecular mass of the A-component $m^A$ is 4 times of that of the B-component $m^B$. Due to the conservation of mass of each component and the total energy of the mixture, the initial condition~\eqref{initial} leads to the following equilibrium states: 
\begin{equation}\label{unequal_eq}
f^\imath(t=\infty,\textbf{v})=\left\{
\frac{1}{Z^\imath}\exp\left(\frac{|\textbf{v}|^2}{T}
\right)-\theta_0\right\}^{-1},
\end{equation} 
where the fugacities $Z^A$ and $Z^B$ of each component and the temperature $T$ of the mixture are $(Z^A,Z^B,T)=(7.9246, 0.7284, 1.1634)$, $(1.3320, 0.3330, 1.9118)$, and $(0.6461, 0.2287, 2.4516)$ for Fermi, classical, and Bose gases, respectively.

In the numerical simulation, the velocity space $[-L,L)^2$ with $L=12$ is discretized by $64\times64$ uniformly-distributed grid points: we choose $L=12$ because the component B has a smaller molecular mass, so it requires larger velocity domain. For the component B, however, $N=64$ and $L=12$ is roughly equivalent to $N=32$ and $L=6$ in the equal-mass mixture in Sec.~\ref{equal_mass}. We also choose $M=5$ and $M_2=10$, as doubling the two values only slightly improves the accuracy. These parameters should have the same order of error as the case of $N=32$ in the equal-mass case considered in Sec.~\ref{equal_mass}, where the conservation of mass and total energy is preserved with the relative error less than $10^{-8}$ and $10^{-6}$, respectively. 

\begin{figure}[h]
	\centering
	\includegraphics[scale=0.6,viewport=40 60 750 470,clip=true]{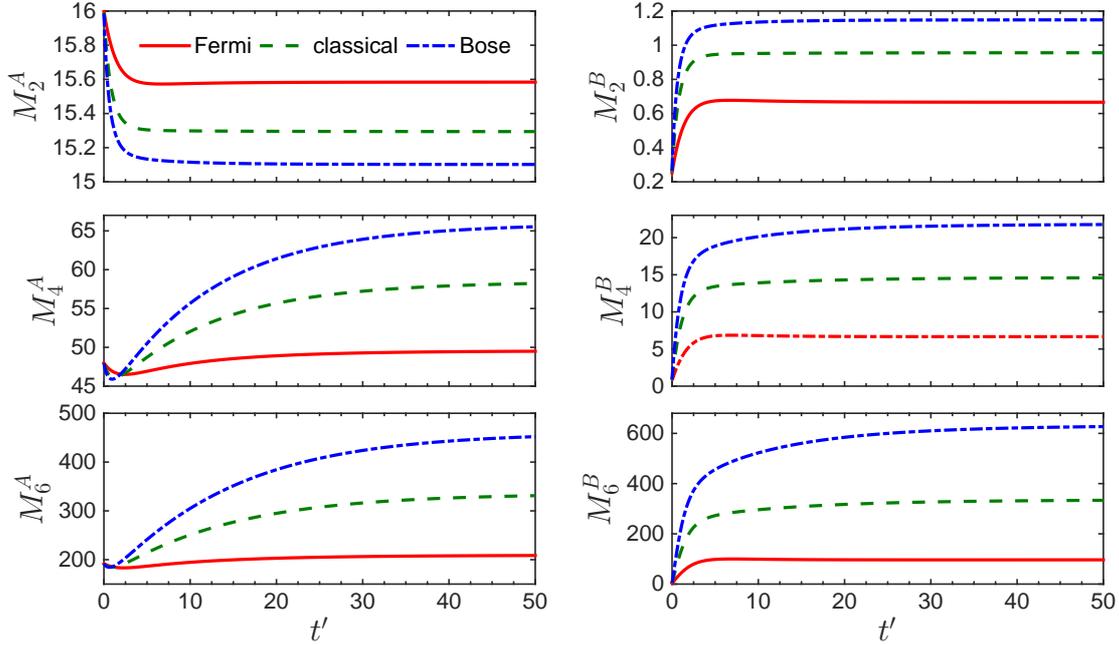}
	\caption{ The relaxation of the second-, fourth-, and sixth-order moments of the VDFs in the binary mixture with $m^A=4m^B$.  Here the moments are defined as $M_i^\imath(t)=(m^\imath/m^A)^3\int\int f^\imath(\textbf{v},t)|\textbf{v}|^id\textbf{v}$, where $i=2,4$, and 6. Other parameters are the same as used in Fig.~\ref{Relax_UnequalMass}. }
	\label{M2M4M6_UnequalMass}
\end{figure}

The relaxation of the two initial VDFs~\eqref{initial} is depicted in Fig.~\ref{Relax_UnequalMass}, while the time evolution of the second-, fourth-, and sixth-order moments are shown in Fig.~\ref{M2M4M6_UnequalMass}. It is seen that near the region $v_1=0$, the VDF of the component A increases monotonically with the time, while that of the component B first increases rapidly, and then decreases as the time $t'$ goes by. This is due to the energy exchange between the two components: from the first row in Fig.~\ref{M2M4M6_UnequalMass} we see that the component B receives the energy from the component A, so the width of the VDF of the component B has to increase while the value of VDF near $v_1=0$ has to decrease. When $t'$ is large enough, the final equilibrium states have been achieved for both components, and the VDFs agree well with the analytical solutions~\eqref{unequal_eq}. Finally, when compared to the equal-mass mixture case without energy transfer between the two components, it is seen in Fig.~\ref{M2M4M6_UnequalMass} that the fourth- and sixth-order moments of the component A first decreases slightly, due to the energy output to the component B, and then increase with the time, while those of the component B always increase until reach the corresponding equilibrium values.

It should be emphasized that the two numerical examples presented in this section only show the correctness of the relaxation to the final equilibrium states. However, whether the relaxation process (i.e. the speed of relaxation) is accurately captured by FSM or not is not clear, since we have no analytical solutions to compare with for quantum gases, although for classical Boltzmann equation of Maxwell molecules (i.e. the intermolecular force is proportional to $r^{-5}$, where $r$ is the intermolecular distance), the relaxation process has been verified by analytical solutions~\cite{lei,Lei2015FSM}. In the next section, we will assess the accuracy of the FSM by comparing the numerical results of  transport coefficients of the quantum Boltzmann equations with analytical and numerical solutions presented in literature~\cite{Watabe2010,bruun_2012,arxiv_sch}.

\section{Transport coefficients}\label{transportCoe}

Compared to classical gases, transport coefficients of quantum gases are hard to measure experimentally. Therefore, an accurate and efficient method is urgently needed to solve the quantum Boltzmann equation. The transport coefficients such as the shear viscosity, thermal conductivity, and diffusion of the quantum gas can be calculated by means of the Chapman-Enskog expansion~\cite{CE}. The basic idea of this expansion is to expand the VDF around the local equilibrium~\eqref{quantum_equilibrium} in terms of a small parameter related to the Knudsen number, which gives the Euler equations at the zeroth-order. For the first-order approximation (i.e. a solution of Eq.~\eqref{Boltzmann_b} in the form of Eq.~\eqref{perturbation} is sought), the Navier-Stokes equations can be derived, where the small perturbation satisfies (in what follows we focus on two-component mixtures; detailed calculation can be found, e.g. in Ref.~\cite{Watabe2010}):
\begin{equation}\label{watabe_transport}
\begin{aligned}[b]
\mathcal{L}^{\imath\jmath}(h^\imath,h^\jmath)=\bigg\{
&\frac{m^\imath}{k_BT}\sum_{ij}D^\imath_{ij}\left[v_{r,i}v_{r,i}-\frac{\delta_{ij}}{d_v}|\textbf{v}_r|^2\right]+\textbf{v}_r\cdot\textbf{d}^\imath\\
&+
\frac{\textbf{v}_{r}\cdot\nabla_\textbf{x}T}{T}\left[\frac{m^\imath|\textbf{v}_r|^2}{2k_BT}-\frac{d_v+2}{2}\frac{G_{(d_v+2)/2}(Z^\imath)}{G_{d_v/2}(Z^\imath)}\right]
\bigg\}f^\imath_{eq}(1+\theta_0f^\imath_{eq}),
\end{aligned}
\end{equation}
where $D_{ij}=(\partial{V_j}/\partial{x_i}+\partial{V_i}/\partial{x_j})/2$ is the rate-of-strain tensor. Note that the first, second, and third terms on the right-hand size of Eq.~\eqref{watabe_transport} are related to the shear viscosity, diffusion, and thermal conductivity, respectively. Since the definition of the coefficient of mass diffusion refers to a state of the gas in which no external forces act on the molecules, and the pressure and temperature of the gas are uniform~\cite{CE}, the complicated expression for $\textbf{d}^\imath$ in Ref.~\cite{Watabe2010} is simplified to 
$
\textbf{d}^\imath=\frac{\nabla_\textbf{x}Z^\imath}{Z^\imath}=\frac{\nabla_\textbf{x}\mu^\imath}{k_BT}$.

The constitutive relations at the first-order Chapman-Enskog expansion are given by		
\begin{equation}
\begin{aligned}[b]
P=\sum_{\imath}\delta_{ij}P^\imath_{\imath\jmath}-2\eta\left[D_{ij}-\frac{\text{Tr}(D_{ij})}{d_v}\delta_{ij}\right],\quad
\textbf{Q}=-\kappa\nabla{T},\quad
\textbf{J}_M=-D\nabla{M},
\end{aligned}
\end{equation}
where $P$ is the total pressure of the mixture, and $\textbf{J}_M$ is the mass current induced by the population difference $M=n^\imath-n^\jmath$. The shear viscosity $\eta$, thermal conductivity $\kappa$, and mass diffusion coefficient $D$ can be found by substituting Eq.~\eqref{perturbation} into Eq.~\eqref{macroscopic_quantities}, where for the shear viscosity, thermal conductivity, and mass diffusion coefficient the perturbation $h$ respectively satisfies the following equations: 
\begin{eqnarray}
\mathcal{L}^{\imath\jmath}(h^\imath,h^\jmath)&=&f^\imath_{eq}(1+\theta_0f^\imath_{eq})\frac{m^\imath}{k_BT}D^\imath_{ij}\left[v_{r,i}v_{r,i}-\frac{\delta_{ij}}{d_v}|\textbf{v}_r|^2\right], \label{linearized1}\\
\mathcal{L}^{\imath\jmath}(h^\imath,h^\jmath)&=&f^\imath_{eq}(1+\theta_0f^\imath_{eq})\frac{\textbf{v}_{r}\cdot\nabla{}T}{T}\left[\frac{m^\imath|\textbf{v}_r|^2}{2k_BT}-\frac{d_v+2}{2}\frac{G_{(d_v+2)/2}(Z^\imath)}{G_{d_v/2}(Z^\imath)}\right], \label{linearized2}\\
\mathcal{L}^{\imath\jmath}(h^\imath,h^\jmath)&=&f^\imath_{eq}(1+\theta_0f^\imath_{eq})\frac{\textbf{v}_r\cdot\nabla_\textbf{x}\mu^\imath}{k_BT}. \label{linearized3}
\end{eqnarray}
For simplicity, in the following calculations, we define terms on the right-hand sides of Eqs.~\eqref{linearized1}-\eqref{linearized3} as the source terms $\mathcal{S}^\imath$.

\subsection{Variational principles}

The complicated mathematical structure of the linearized Boltzmann collision operator $\mathcal{L}^{\imath\jmath}$ makes the exact solution for the perturbation $h$ in Eqs.~\eqref{linearized1}-\eqref{linearized3} extremely difficult to find. Therefore,  variational principles are used to find the upper and lower bound of the transport coefficient~\cite{Smith_book}. A simple way is to use the following ansatz:
\begin{equation}\label{ansatz}
h^\imath=C^\imath \mathcal{S}^\imath, \quad\quad \imath=A, B,
\end{equation} 
where $C^\imath$ are constants, whose values can be obtained by solving the following two linear equations of $C^A$ and $C^B$:
\begin{equation}\label{variation_ana}
\int \mathcal{L}^{\imath\jmath}(C^\imath\mathcal{S}^\imath,C^\jmath\mathcal{S}^\jmath) \frac{\mathcal{S}^\imath}{f^\imath_{eq}(1+\theta_0f^\imath_{eq})}d\textbf{v}=\int  \frac{(\mathcal{S}^\imath)^2}{f^\imath_{eq}(1+\theta_0f^\imath_{eq})}d\textbf{v}, \quad\quad \imath=A, B.
\end{equation}

Expressions for the two constants $C^A$ and $C^B$ can be simplified analytically, and then solved by numerical quadrature (for the classical Boltzmann equation with some special forms of differential cross-section, analytical solution may be derived), see Eq.~\eqref{eta_r} below. Also, it can be computed by the FSM developed in this paper.

The variational principle~\eqref{ansatz} predicts the lower bound of transport coefficients. For the classical Boltzmann equation, this variational principle gives accurate transport coefficients for Maxwell molecules, while for hard-sphere molecules it underpredicts the transport coefficients by only about 2 percent~\cite{CE}. Whether this conclusion holds for quantum gases or not is not clear; this will be assessed in the following numerical examples.

\subsection{Direct numerical simulation}

A direct numerical solution of the linear equations in Eqs.~\eqref{linearized1}-\eqref{linearized3} is necessary to find accurate transport coefficients. To this end, we first define the following two constants as the maximum values of the equilibrium collision frequencies in Eq.~\eqref{equilibrium_frequency}, for classical gases:
$
\mu^\imath=\sum_{\jmath}\mu_c^{\imath\jmath}(\textbf{v}=0)$ with $\imath=A, B$. 
Then, the linear perturbation can be solved through the following iterative scheme~\cite{Lei_POF2015}:
\begin{equation}\label{iteration_transport}
\begin{aligned}[b]
h^{\imath,j+1}=\frac{-\mathcal{S}^\imath+\mathcal{L}^{\imath\jmath}(h^{\imath,j},h^{\jmath,j})+\mu^\imath{h}^\imath}{\mu^{\imath}}, \quad\quad \imath=A, B,
\end{aligned}
\end{equation}
where the subscript $j$ and $j+1$ are the iteration steps.

The reason to use $\mu^\imath$ in the denominator of Eq.~\eqref{iteration_transport} instead of the equilibrium collision frequency $\mu_{c}^{\imath\jmath}$, as normally used in the iterative scheme~\cite{Lei_POF2015}, is that the collision frequency approximated by FSM approaches zero at large relative collision velocity $\textbf{u}$ for the special differential cross-section~\eqref{two_d_s}. Therefore, the iteration will diverge when $\mu_{c}^{\imath\jmath}$ is used in the denominator. Numerical simulations below have proven that the iterative scheme~\eqref{iteration_transport} is unconditionally stable, while using $\mu_{c}^{\imath\jmath}$ in the denominator results in no converged solution when the quantum gas is highly degenerated, that is, when the fugacity $Z$ approaches infinity for Fermi gas and $Z$ approaches one for Bose gas.

In the following numerical simulations, the iteration is terminated until the relative error in the transport coefficient between two consecutive steps is less than $10^{-5}$. Starting from the zero perturbation, normally only several dozen iterations are needed to satisfy this convergence criterion.

\subsection{Results: three-dimensional case}\label{D3case}

We consider the two-component population balanced Fermi gases, with $m^A=m^B=m$. In the most experiments, the two components move together and only one VDF is enough to describe the system state. Due to Pauli's exclusion principle, the $s$-wave scattering happens between molecules with different spins. As a consequent, only the cross-collision operators are considered. For simplicity, the hard-sphere molecular model is used, where the differential cross-section is
${d\sigma^{\imath\jmath}}/{d\Omega}=a_s^2$.

Applying the Chapman-Enskog expansion to the quantum Boltzmann equation, one obtains 
the shear viscosity and thermal conductivity as~\cite{Watabe2010}
\begin{equation}\label{eta_r} 
\begin{aligned}[b]
\eta=\frac{5m}{32a_s^2I_B}\sqrt{\frac{k_B
		T}{m}}{G}^2_{5/2}(Z), \quad
\kappa=\frac{75k_B}{256a_s^2I_A}\sqrt{\frac{k_B
		T}{m}}\left[\frac{7}{2}{{G}_{7/2}(Z)}
-\frac{5}{2}\frac{{G}^2_{5/2}(Z)}{{G}_{3/2}(Z)}\right]^2, 
\end{aligned}
\end{equation}
where 
\begin{eqnarray*}
I_A&=&\int_0^\infty d\xi_0\xi_0^4\int_0^\infty
d\xi'{\xi'^7}\int_0^1dy'\int_0^1dy''F\cdot(y'^2+y''^2-2y'^2y''^2), \\
I_B&=&\int_0^\infty d\xi_0\xi_0^2\int_0^\infty
d\xi'{\xi'^7}\int_0^1dy'\int_0^1dy''F\cdot(1+y'^2+y''^2-3y'^2y''^2),\\
F&=&\frac{Z^2\exp(-\xi_0^2-\xi'^2)}{[1-\theta_0Z\exp(-\xi_1^2)][1-\theta_0Z\exp(-\xi_2^2)][1-\theta_0Z\exp(-\xi_3^2)][1-\theta_0Z\exp(-\xi_4^2)]},
\end{eqnarray*}
$\xi_1^2=(\xi_0^2+2\xi_0\xi'y'+\xi'^2)/2$, 
$\xi_2^2=(\xi_0^2-2\xi_0\xi'y'+\xi'^2)/2$, 
$\xi_3^2=(\xi_0^2+2\xi_0\xi'y''+\xi'^2)/2$, and 
$\xi_4^2=(\xi_0^2-2\xi_0\xi'y''+\xi'^2)/2$.

\begin{figure}[t]
	\centering
	\includegraphics[scale=0.6,viewport=10 0 560 410,clip=true]{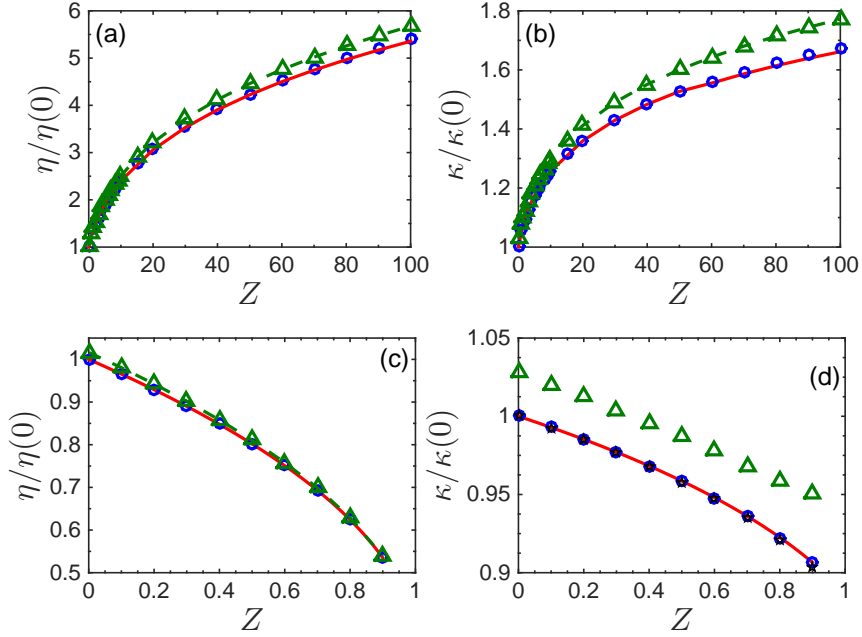}
	\caption{The shear viscosity $\eta$ and thermal conductivity $\kappa$ of Fermi (top row) and Bose (bottom row) gases, as a function of the fugacity $Z$,  where $\eta_0$ and $\kappa_0$ are respectively the shear viscosity and thermal conductivity at the classical limit $Z=0$, which are obtained from the analytical solution~\eqref{eta_r} that is derived from the variational principle~\cite{Nikuni1998,Watabe2010}. Solid lines: analytical solution~\eqref{eta_r}. Circles: numerical solutions using the variational principle, i.e. by solving Eq.~\eqref{variation_ana} numerically via  FSM. Triangles: numerical results obtained by solving Eq.~\eqref{iteration_transport} via FSM. }\label{compare_transport_3d}
\end{figure}

For the one-component Bose gas, the differential cross-section is ${d\sigma^{\imath\jmath}}/{d\Omega}=2a_s^2$~\cite{Nikuni1998}, so the shear viscosity and thermal conductivity will be four times smaller than those of the population balanced Fermi gas.

Figure~\ref{compare_transport_3d} shows the shear viscosity and thermal conductivity of the quantum Fermi and Bose gases as a function of the fugacity. It is seen that the shear viscosity and thermal conductivity of the Fermi (Bose) gas increase (decrease) with the fugacity $Z$.  FSM solutions of the variational equation~\eqref{variation_ana} agree well with the analytical solutions~\eqref{eta_r} obtained by the same variational principle, which proves that our FSM has a high accuracy.

With the accuracy of the FSM verified by  analytical solutions, we assess the accuracy of the variational principle that only gives the lower bound of the transport coefficient, by solving the linearized equation using the iterative method~\eqref{iteration_transport}. Results are shown in Fig.~\ref{compare_transport_3d} as triangles. For Fermi gas, at $Z$ increases from 0 to 100, the relative error between the accurate shear viscosity (thermal conductivity) and those from the variational principle increases from 1.6\% (2.8\%) to 5.2\% (6\%).  For Bose gas, this relative error in thermal conductivity increases  from about 2.8\% when $Z=0$ to 5.2\% when $Z=0.9$, while that in shear viscosity decreases from 1.6\% when $Z=0$ to 0.2\% when $Z=0.9$.

\begin{figure}[t]
	\centering
	\includegraphics[scale=0.43,viewport=10 0 500 370,clip=true]{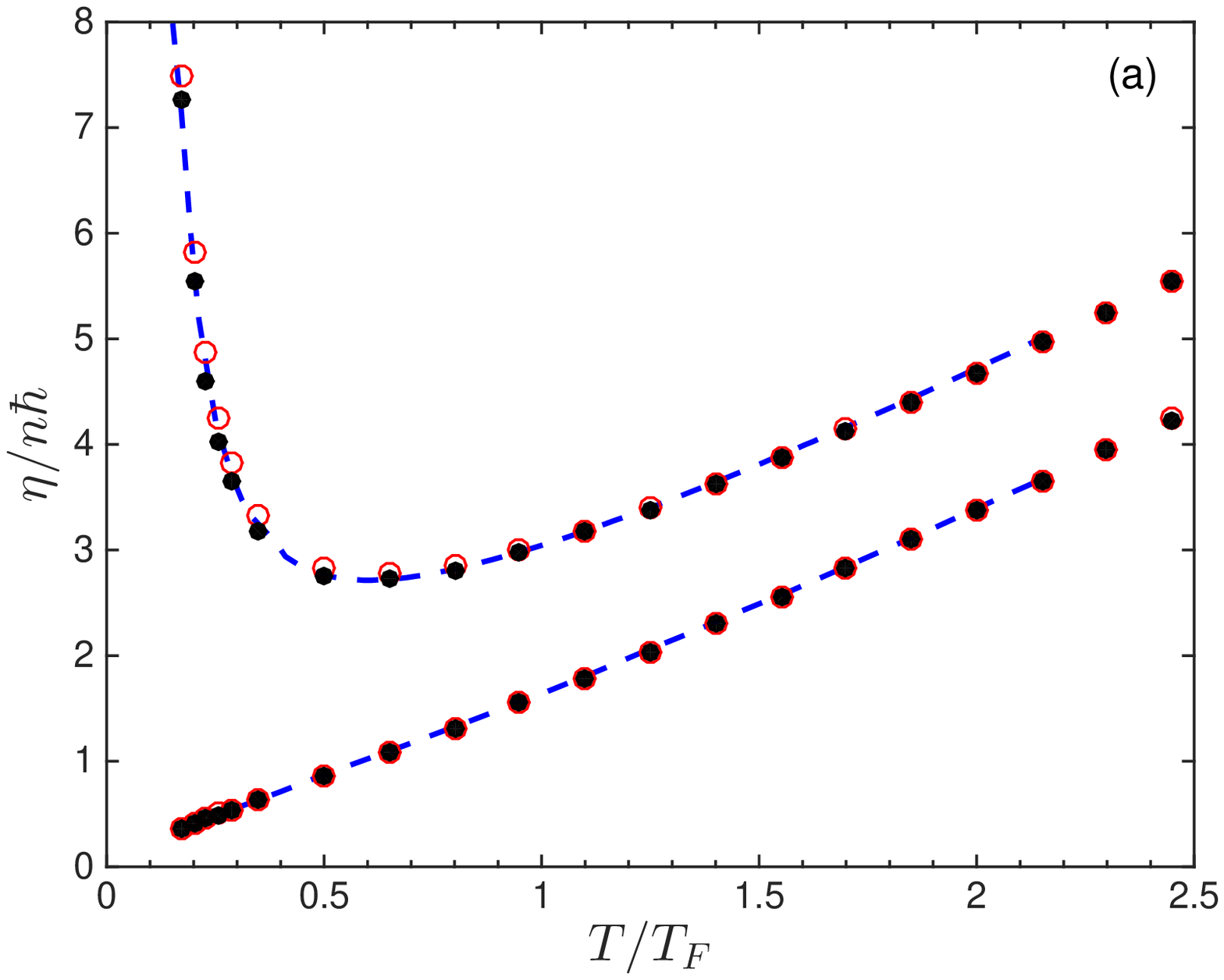}
	\hskip 0.5cm
	\includegraphics[scale=0.43,viewport=10 0 500 370,clip=true]{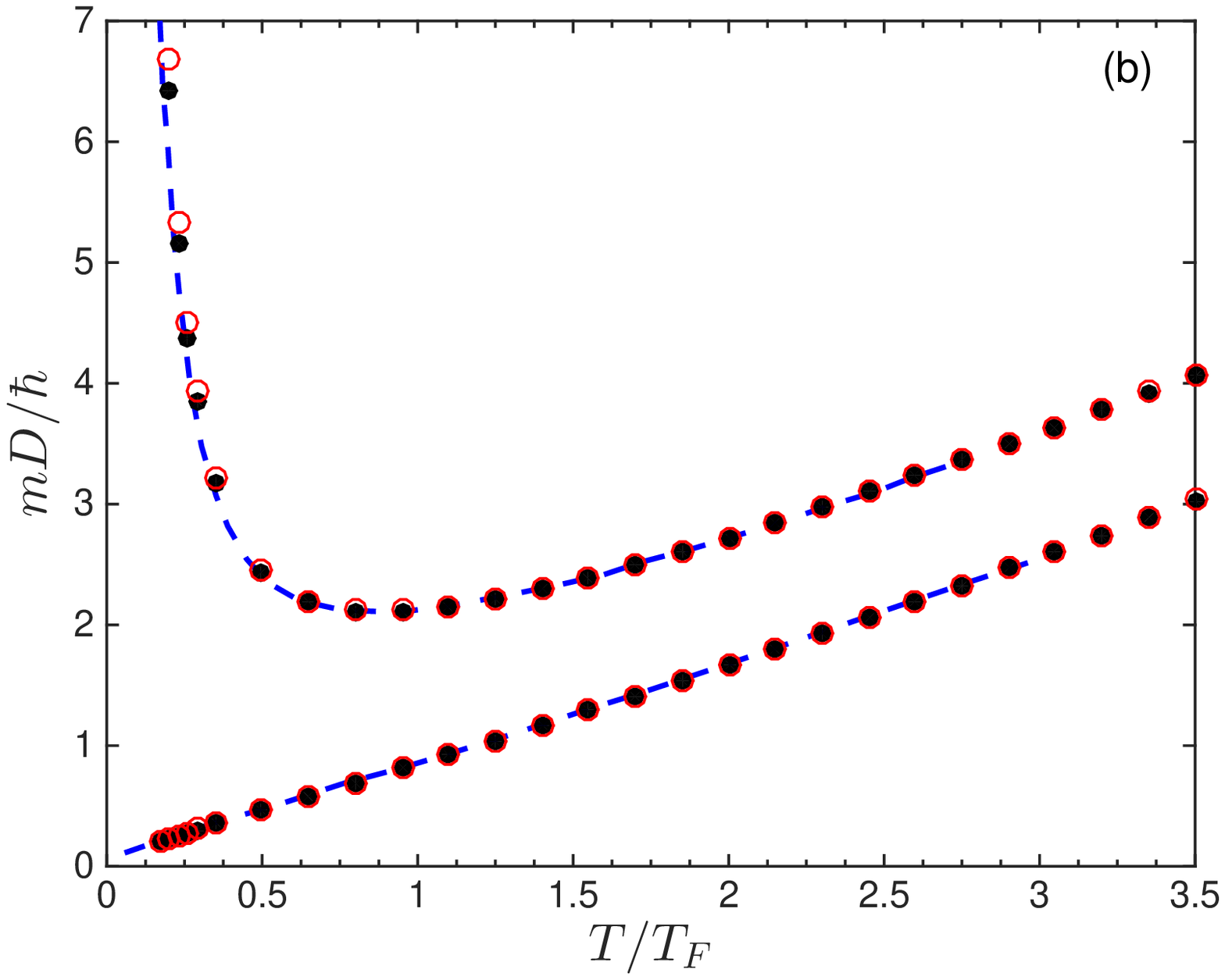}\\
	\vskip 0.5cm
		\includegraphics[scale=0.43,viewport=10 0 500 370,clip=true]{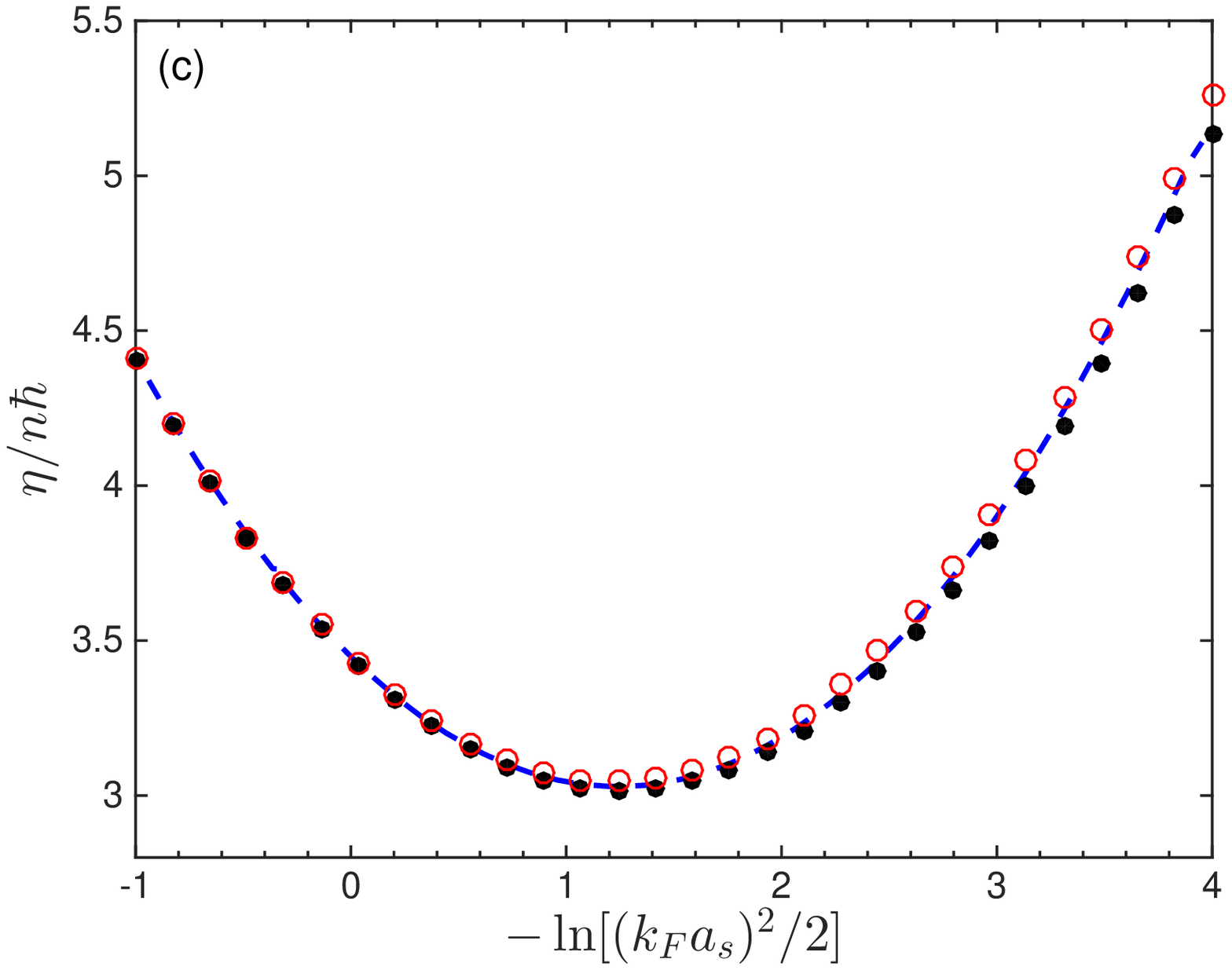}
	\hskip 0.5cm
	\includegraphics[scale=0.43,viewport=10 0 500 370,clip=true]{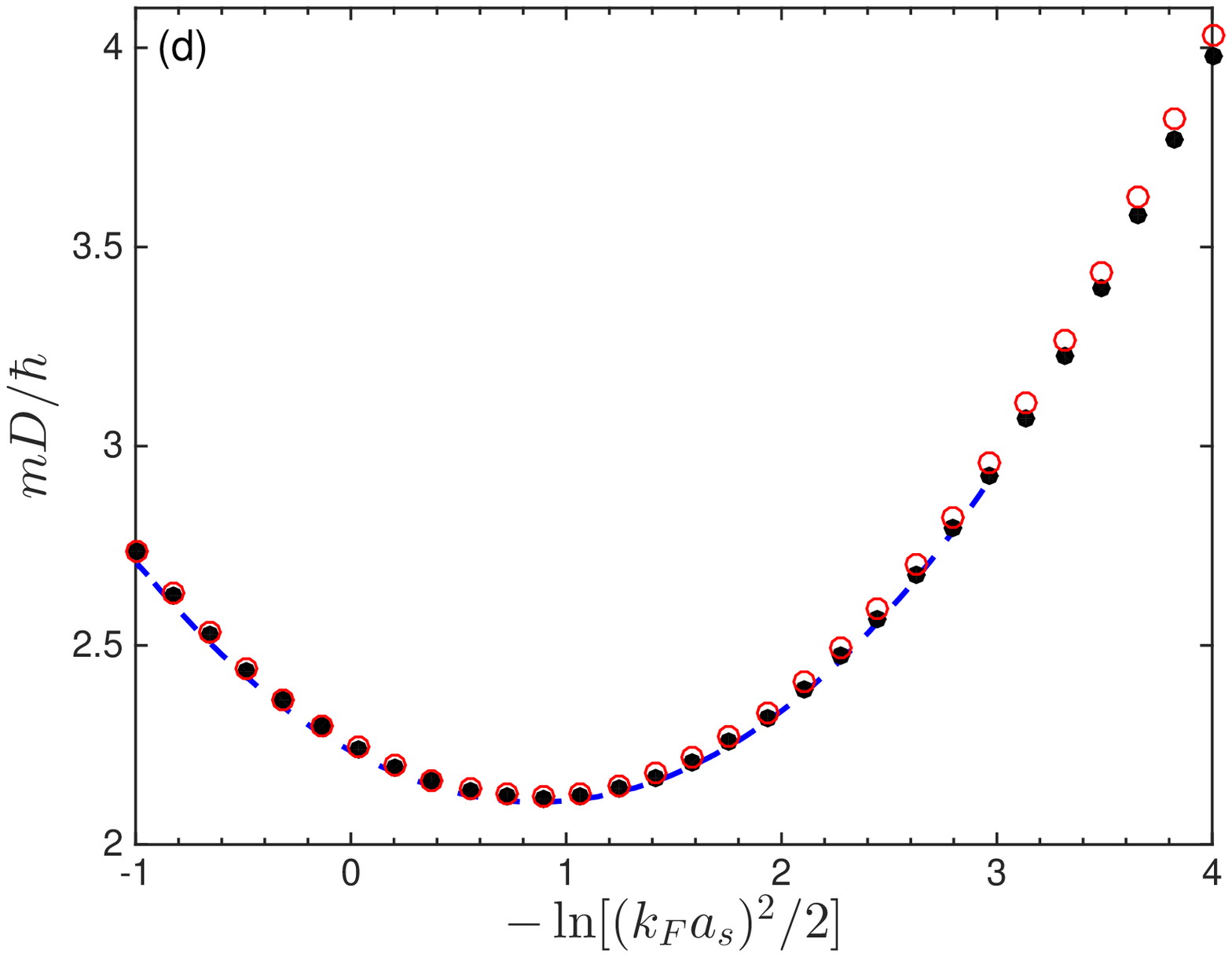}
	\caption{The normalized shear viscosity (a, c) and mass diffusion coefficient (b, d) of the 2D Fermi gas as  functions of the (a, b) normalized temperature $T/T_F$ at $(k_Fa_s)^2={2\exp(-1)}$ and (c, d) s-wave scattering length $a_s$ at $T/T_F=1$. Dashed lines represent results from the variation principle adopted from Ref.~\cite{bruun_2012}. Solid circles: FSM solutions of the variational principle~\eqref{variation_ana}. Open circles: FSM solutions of the iterative scheme~\eqref{iteration_transport}. Nearly-straight lines in (a) and (b) are the corresponding results for classical gases. Note that $T_F=(\hbar{k_F})^2/2mk_B$ is the Fermi temperature, and  $k_F=\sqrt{2\pi{n}}$ is the Fermi wave vector, with $n$ being the total number density of both spin components.}\label{shear_2d}
\end{figure}

\subsubsection{Shear viscosity of the mass-balanced mixture}

We first consider the equal-mass mixture, i.e. $m^A=m^B=m$. Numerical results for the shear viscosity and spin diffusion coefficients are shown in Fig.~\ref{shear_2d}, for a wide range of the temperature and s-wave scattering length. It is clear that the variational solutions solved by FSM agree well with the numerical solutions of Brunn~\cite{bruun_2012} for both classical and Fermi gases, while the accurate shear viscosity and mass diffusion coefficient obtained from the iterative scheme~\eqref{iteration_transport} have very limited difference to the variational solutions (i.e. less than 1\%) when $T/T_F<1$. However, at very small values of $T/T_F$, accurate transport coefficients are larger than the variational ones by about 5\% for Fermi gas. This observation is consistent with the 3D Fermi gas case investigated in Sec.~\ref{D3case}. 


%

\begin{figure}[t]
	\centering
	\includegraphics[scale=0.45,viewport=10 0 500 370,clip=true]{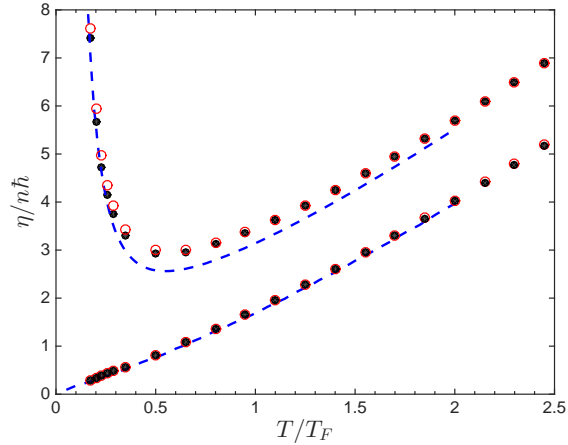}
	\caption{The normalized shear viscosity of the 2D Fermi gas as a function of the temperature, where the interaction strength between fermions with equal mass but opposite spins is $(k_Fa_s)^2={2}$. Dashed lines represent results from the variation principle adopted from Ref.~\cite{arxiv_sch}. Solid circles:  FSM solutions of the variational principle~\eqref{variation_ana}. Open circles: FSM solutions of the iterative scheme~\eqref{iteration_transport}.}\label{shear_2d_schafer}
\end{figure}

We continue to compare our FSM solutions 
to the numerical solutions by provided by Sch\"{a}fer~\cite{arxiv_sch} in Fig.~\ref{shear_2d_schafer}. The agreement is acceptable in general, especially for the case of classical gases. For Fermi gases, the shear viscosity obtained from FSM agrees well with the variational solutions~\cite{arxiv_sch} in the low and high temperature limits. However, in the intermediate regime (near $T/T_F=0.5$) where the shear viscosity is minimum, both of our FSM solutions, obtained from the variational principle~\eqref{variation_ana} and the iterative scheme~\eqref{iteration_transport}, are higher than the variational results of Sch\"{a}fer~\cite{arxiv_sch} by about 15\%.

\subsubsection{Shear viscosity of mass-imbalanced mixtures}

\begin{figure}[t]
	\centering
	\includegraphics[scale=0.38,viewport=10 0 560 410,clip=true]{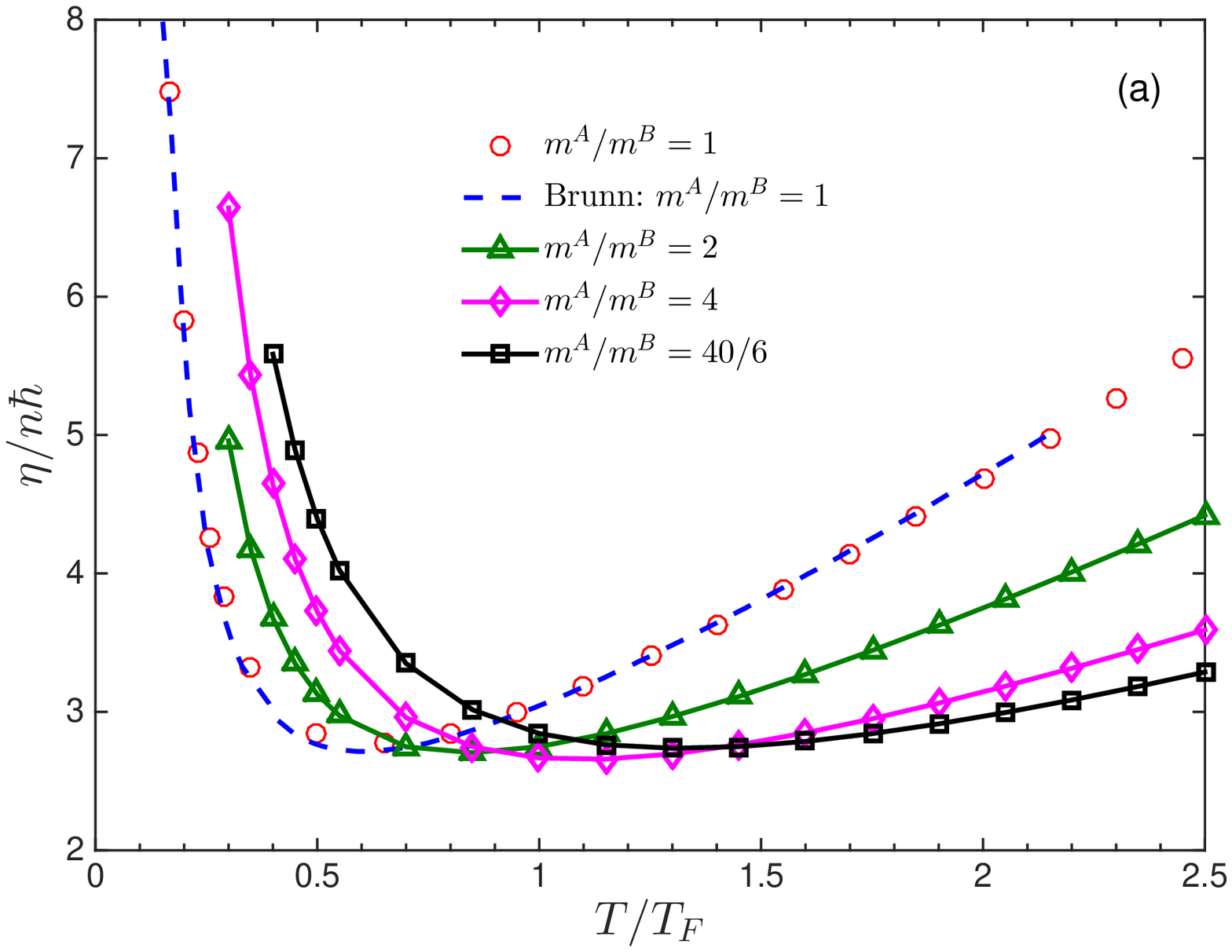}
	\hskip 0.5cm
	\includegraphics[scale=0.38,viewport=10 0 560 410,clip=true]{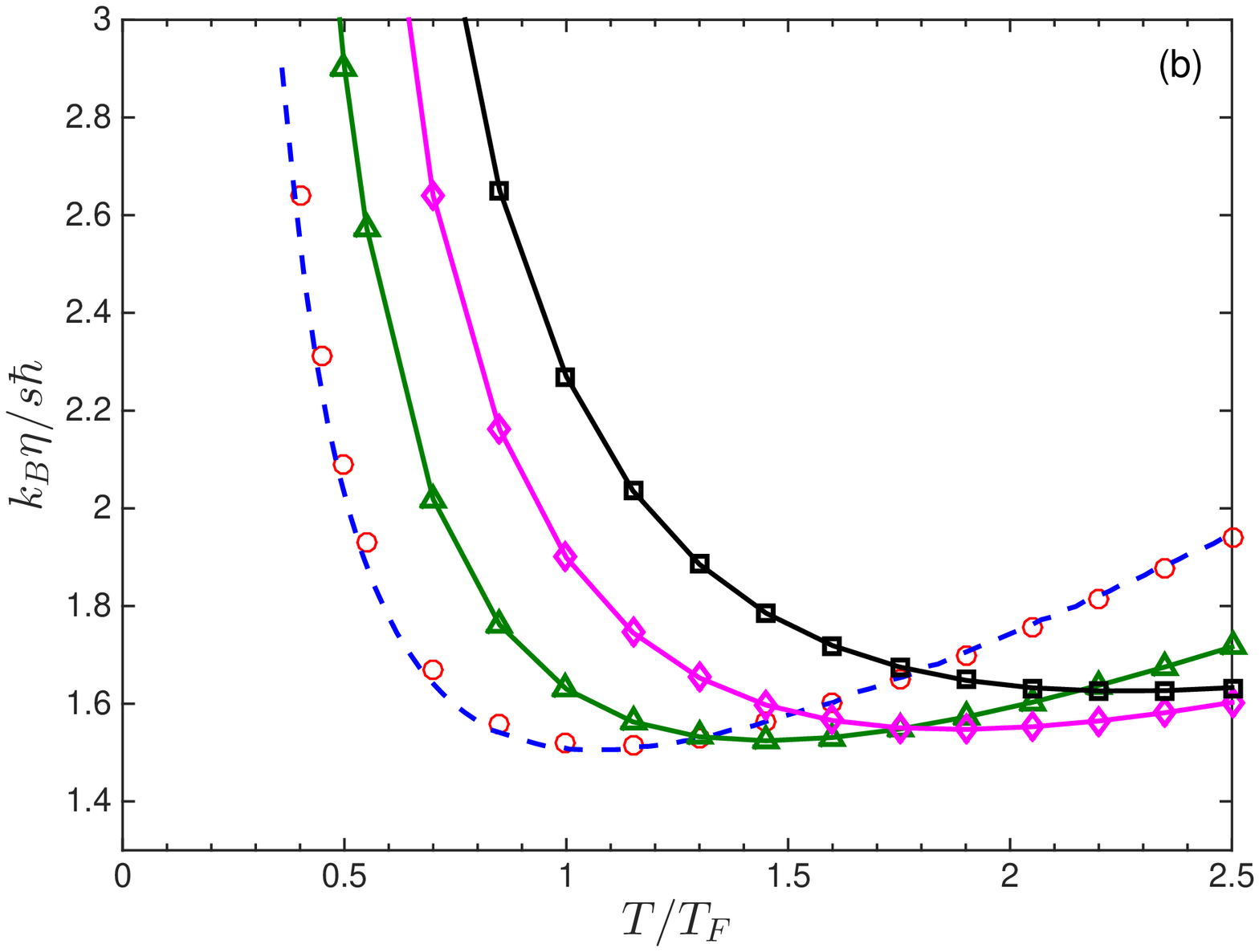}\\
	\vskip 0.5cm
	\includegraphics[scale=0.38,viewport=10 0 560 410,clip=true]{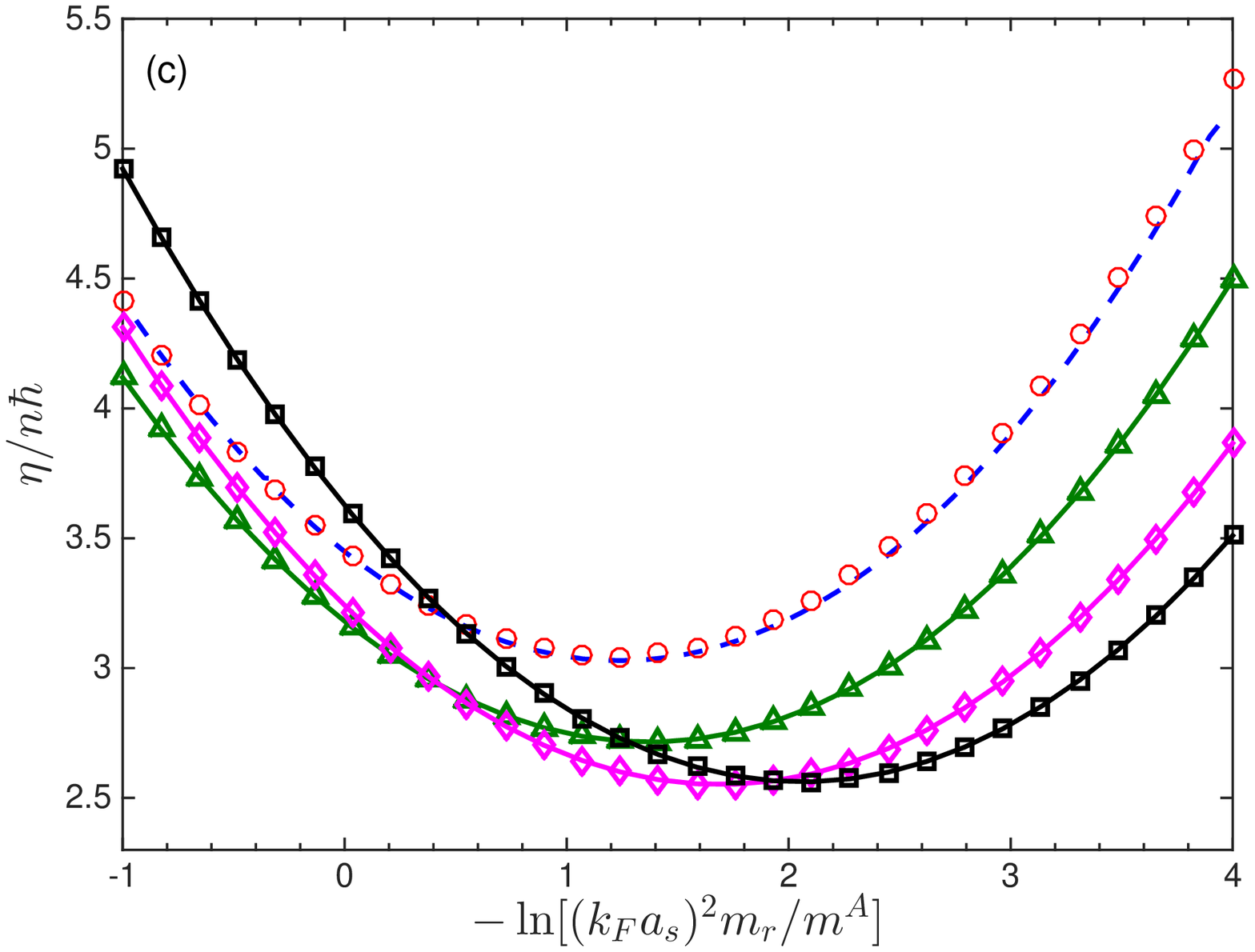}
	\hskip 0.5cm
	\includegraphics[scale=0.38,viewport=10 0 560 410,clip=true]{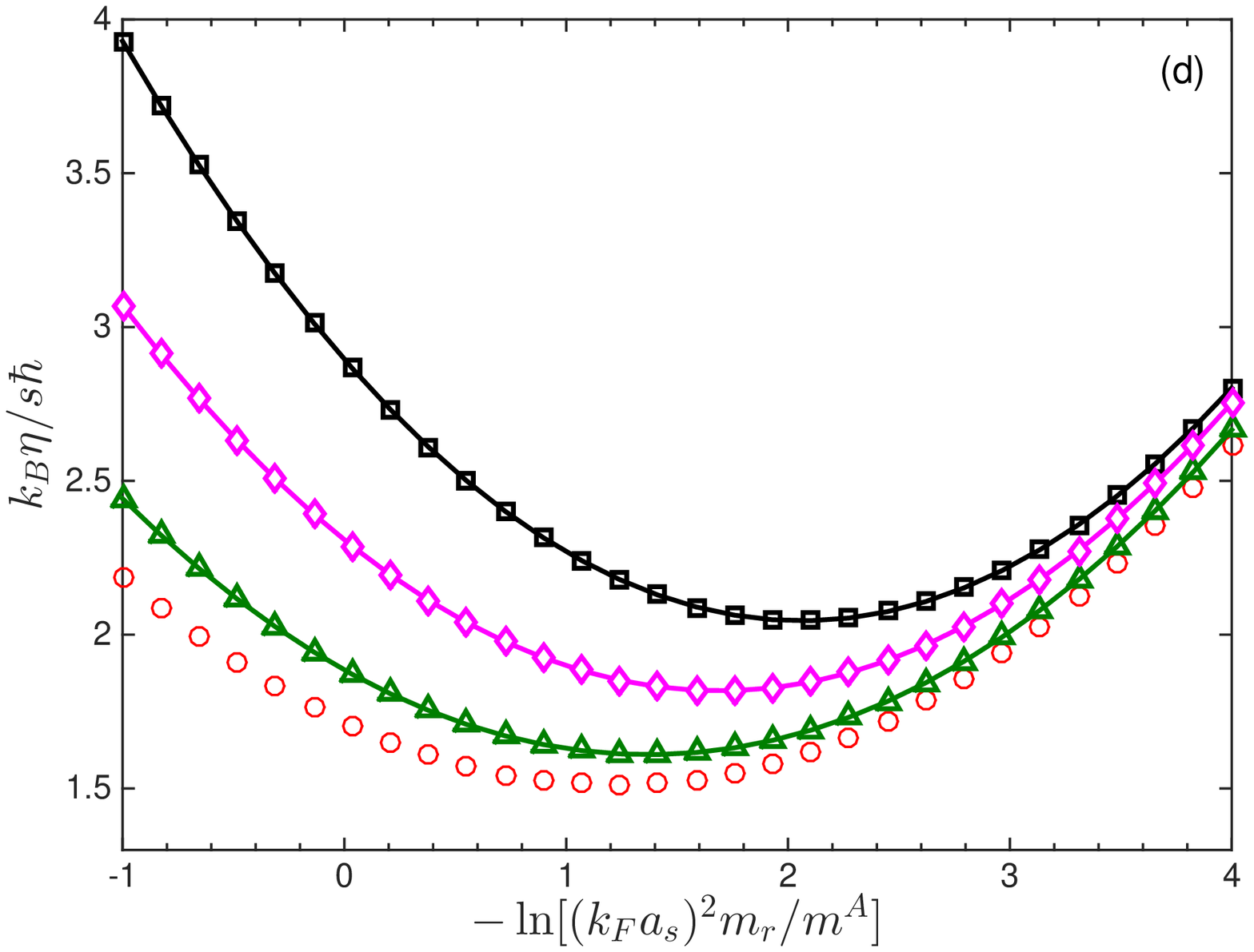}
	\caption{The shear viscosity of equal-mole mixture of 2D quantum Fermi gas, where the molecular mass of each components are different. 
	The shear viscosity (a) and viscosity-entropy ratio (b) of the 2D Fermi gas as a function of the  normalized temperature $T/T_F$ at $(k_Fa_sm_r/m^A)^2={\exp(-1)}$. The shear viscosity (c) and viscosity-entropy density ratio (d) of the 2D Fermi gas as a function of the s-wave scattering length $(k_Fa_sm_r/m^A)^2$ when $T/T_F=1$. Symbols:  FSM solutions of the iterative scheme~\eqref{iteration_transport}. Note that  $T_F=(\hbar{k_F})^2/2m^Ak_B$ is the Fermi temperature of the A-component, and  $k_F=\sqrt{2\pi{n}}$ is the Fermi wave vector, with $n$ being the total number density of both spin components.}\label{PRA_Bruun}
\end{figure}

We further calculate the shear viscosity of the equal-mole mixture of 2D Fermi gas, where the A-component has a larger molecular mass than the B-component.  Fig.~\ref{PRA_Bruun} plots the shear viscosity when $m^A/m^B=1$, 2, 4, and $40/6$. It is observed in Fig.~\ref{PRA_Bruun}(a) that, when the s-wave scattering length is fixed, that is, when the ratio of the two-body binding energy $E_b=1/2m_ra_s^2$ to the Fermi energy of the A-component is equal to $\exp(1)$, the shear viscosity first decreases when the temperature increases, and then increases with the temperature, for all the molecular mass ratios considered. However, the reduced temperature $T/T_F$ at which the minimum shear viscosity is reached increases with the mass ratio. The same trend applies also to the viscosity-entropy density ratio in Fig.~\ref{PRA_Bruun}(b). Interestingly, in Fig.~\ref{PRA_Bruun}(a) we see that the minimum shear viscosity almost remains unchanged when the molecular mass ratio varies; this is in sharp contrast to the variational results~\cite{bruun_2012}, which states that the shear viscosity should be proportional to the reduced mass, i.e. should decrease when the mass ratio increases. This discrepancy may be caused by the fact that the variational ansatz used in Eq.~(4) of Ref.~\cite{bruun_2012} is different to ours in Eq.~\eqref{ansatz} when the molecular mass ratio is not one.

Figure~\ref{PRA_Bruun}(c) shows the variation of the shear viscosity as the interaction strength, when the temperature of the mixture is equal to the Fermi temperature of the A-component. When the molecular mass ratio is fixed, there is a minimum value of shear viscosity; and it seems that this minimum viscosity decreases when the mass ratio increases, but quickly saturated at $m^A/m^B=40/6$. In addition, at small enough interaction strength, i.e. in the right part of Fig.~\ref{PRA_Bruun}(c), the shear viscosity decreases when the molecular mass ratio increases, while at large interaction strength, there is no monotonous relation between the shear viscosity and mass ratio.

Figure~\ref{PRA_Bruun}(b) and (d) depict the ratio between the shear viscosity and entropy density. It is clear that the minimum viscosity-entropy ratio does not change much when the molecular mass ratio varies. Although Brunn~\cite{bruun_2012} claimed that the universal bound of the viscosity-entropy density ratio obtained from string theory methods~\cite{viscosity_entropy}
\begin{equation}\label{viscosity_entropy}
\frac{k_B\eta}{s\hbar}>\frac{1}{4\pi}
\end{equation} 
may be violated at large molecular mass ratios, our numerical calculations suggested this is not the case, at least for the quantum Boltzmann equation with the differential cross-section~\eqref{two_d_s}.

\section{Conclusions}\label{ConFuture}

We have developed a FSM to solve the quantum Boltzmann equation for gas mixtures with general forms of differential cross-sections, with the computational cost of the FSM proposed in this paper is $O(M^{d_v-1}M_2N^{d_v}\log{N})$, which is the same for the Boltzmann collision operator when the general form of intermolecular potential is considered~\cite{Lei_POF2015}. The spatially-homogeneous relaxation problem has been used to determine factors that affect the accuracy of the FSM. It has been shown that, the solid angle (or polar angle in the two-dimensional problem) can be discretized uniformly by $M^2=5\times5$ (or $M=5$) points, while the number of abscissas in Gauss-Legendre quadrature used in Eq.~\eqref{kernel_model_GL} can be as small as $M_2=10$, when $N=32$ velocity points are used to discretize the velocity distribution function in each direction.  The FSM handles the collision in the frequency space, and conserves the mass exactly, while the momentum and energy are conserved with spectral accuracy, provided that the discretized frequency space is wide enough to cover the whole spectrum of the velocity distribution function.

Based on the variational principle that predicts the lower bounds of transport coefficients, the shear viscosity and thermal conductivity have been calculated by the FSM for both quantum Fermi and Bose gases.  Comparisons with the analytical solutions demonstrated the accuracy of the proposed FSM. Accurate transport coefficients are obtained by solving the linearized Boltzmann collision operator via the iterative scheme~\eqref{iteration_transport}. As expected, these transport coefficients are larger than those from the variational principle. Generally speaking, the relative error between the accurate and variational transport coefficients increases with the fugacity. The shear viscosity of a two-dimensional equal-mole mixture of Fermi gases has also been investigated for components with different molecular masses. Our numerical solutions suggested that the universal bound of the viscosity-entropy density ratio~\eqref{viscosity_entropy} predicted by the string theory is satisfied.

Finally, we pointed that the established accurate FSM to solving the quantum Boltzmann collision operator are ready to be used to calculate the transport coefficients of noble gases based on the \textit{ab initio} potentials~\cite{Song2010Transport,Sharipov2017JCP}. Also, the FSM can be used to assess the accuracy of quantum kinetic models~\cite{Wu2012modelling,Yang2013ES,Diaz2016JCP}. Furthermore, the FSM  can be incorporated into other multi-scale methods~\cite{Changliu2016,XiaoTB2017JCP} that solve the Boltzmann equation accurately and efficiently from the hydrodynamic to free-molecular flow regimes, which is frequently encountered in experiments where the quantum gas is trapped so that its density is maximum at the trap center (i.e. hydrodynamic regime) and vanishes near the trap edge (i.e. free molecular flow regime). In the future we will investigate the interesting spatially-inhomogeneous oscillations~\cite{Vogt2012,WuleiEPL2012,WuleiPRA2012} and spin diffusion~\cite{Somme2011, Sommer2011b, Koschorreck2013} in quantum gases.

\section*{Acknowledgments}

This work is financially supported by the UK's Engineering and Physical Sciences Research Council (EPSRC) under grant EP/R041938/1.


\bibliographystyle{elsarticle-num}
\bibliography{Bib}

\begin{thebibliography}{10}
\expandafter\ifx\csname url\endcsname\relax
  \def\url#1{\texttt{#1}}\fi
\expandafter\ifx\csname urlprefix\endcsname\relax\def\urlprefix{URL }\fi
\expandafter\ifx\csname href\endcsname\relax
  \def\href#1#2{#2} \def\path#1{#1}\fi

\bibitem{RMP1}
F.~Dalfovo, S.~Giorgini, L.~P. Pitaevskii, S.~Stringari, {Theory of
  Bose-Einstein condensation in trapped gases}, Rev. Mod. Phys. 71 (1999) 463.

\bibitem{Anderson1995Science}
M.~H. Anderson, J.~R. Ensher, M.~R. Matthews, C.~E. Wieman, E.~A. Cornell,
  Observation of {Bose-Einstein} condensation in a dilute atomic vapor, Science
  269 (1995) 198--201.

\bibitem{Jin2003Nature}
M.~Greiner, C.~A. Regal, D.~S. Jin, {Emergence of a molecular Bose--Einstein
  condensate from a Fermi gas}, Nature 426 (2003) 537--540.

\bibitem{Uehling1933}
E.~A. Uehling, G.~E. Uhlenbeck, {Transport phenomena in Einstein-Bose and
  Fermi-Dirac gases. I}, Phys. Rev. {43} ({1933}) {0552--0561}.

\bibitem{JacksonPRL2001}
B.~Jackson, E.~Zaremba, {Finite-temperature simulations of the scissors mode in
  Bose-Einstein condensed gases}, Phys. Rev. Lett. 87 (2001) 100404.

\bibitem{JacksonPRA2002}
B.~Jackson, E.~Zaremba, Modeling {Bose-Einstein} condensed gases at finite
  temperature with {N-body} simulations, Phys. Rev. A 66 (2002) 033606.

\bibitem{CE}
S.~Chapman, T.~Cowling, {The Mathematical Theory of Non-uniform Gases},
  Cambridge University Press, 1970.

\bibitem{GarciaPRE2003}
A.~L. Garcia, W.~Wagner, {Direct simulation Monte Carlo method for the
  Uehling-Uhlenbeck-Boltzmann equation}, Phys. Rev. E 68 (2003) 056703.

\bibitem{Bird1994}
G.~A. Bird, Molecular Gas Dynamics and the Direct Simulation of Gas Flows,
  Oxford Science Publications, Oxford University Press Inc, New York, 1994.

\bibitem{BorowikJCP2017}
P.~Borowik, J.~Thobel, L.~Adamowicz, {Modified Monte Carlo method for study of
  electron transport in degenerate electron gas in the presence of
  electron-electron interactions, application to graphene}, J. Comput. Phys.
  341 (2017) 397--405.

\bibitem{Yano2017JCP}
R.~Yano, Fast and accurate calculation of dilute quantum gas using
  {Uehling-Uhlenbeck} model equation, J. Comput. Phys. 330 (2017) 1010--1021.

\bibitem{Vogt2012}
E.~Vogt, M.~Feld, B.~Fr\"{o}hlich, D.~Pertot, M.~Koschorreck, M.~K\"{o}hl,
  Scale invariance and viscosity of a two-dimensional {F}ermi gas, Phys. Rev.
  Lett. 108 (2012) 070404.

\bibitem{WuleiEPL2012}
L.~Wu, Y.~H. Zhang, Numerical investigation of the radial quadrupole and
  scissors modes in trapped gases, Europhys. Lett. 97 (2012) 16003.

\bibitem{WuleiPRA2012}
L.~Wu, Y.~H. Zhang, {Applicability of the Boltzmann equation for a
  two-dimensional Fermi gas}, Phys. Rev. A 85 (2012) 045601.

\bibitem{Mouhot2006}
C.~Mouhot, L.~Pareschi, {Fast algorithms for computing the Boltzmann collision
  operator}, Math. Comput {75} ({2006}) {1833--1852}.

\bibitem{Lei_POF2015}
L.~Wu, H.~H. Liu, Y.~H. Zhang, J.~M. Reese, Influence of intermolecular
  potentials on rarefied gas flows: {Fast spectral solutions of the Boltzmann
  equation}, Phys. Fluids 27 (2015) 082002.

\bibitem{lei}
L.~Wu, C.~White, T.~J. Scanlon, J.~M. Reese, Y.~H. Zhang, {Deterministic
  numerical solutions of the Boltzmann equation using the fast spectral
  method}, J. Comput. Phys. 250 (2013) 27--52.

\bibitem{lei_jfm1}
L.~Wu, J.~M. Reese, Y.~H. Zhang, {Solving the Boltzmann equation by the fast
  spectral method: application to microflows}, J. Fluid Mech. 746 (2014)
  53--84.

\bibitem{Wu2017JCP_iterative}
L.~Wu, J.~Zhang, H.~H. Liu, Y.~H. Zhang, J.~M. Reese, A fast iterative scheme
  for the linearized {Boltzmann equation}, J. Comput. Phys. 338 (2017)
  431--451.

\bibitem{WUjfm2014}
L.~Wu, J.~M. Reese, Y.~H. Zhang, Oscillatory rarefied gas flow inside
  rectangular cavities, J. Fluid Mech. 748 (2014) 350--367.

\bibitem{WuPRE2016}
L.~Wu, Sound propagation through a rarefied gas in rectangular channels, Phys.
  Rev. E 94 (2016) 053110.

\bibitem{Wu2014}
L.~Wu, C.~White, T.~J. Scanlon, J.~M. Reese, Y.~H. Zhang, {A kinetic model of
  the Boltzmann equation for nonvibrating polyatomic gases}, J. Fluid Mech. 763
  (2015) 24--50.

\bibitem{Lei2015FSM}
L.~Wu, J.~Zhang, J.~M. Reese, Y.~H. Zhang, A fast spectral method for the
  {Boltzmann} equaiton for monatomic gas mixtures, J. Comput. Phys. 298 (2015)
  602--621.

\bibitem{MinhTuanHo2016}
M.~T. Ho, L.~Wu, I.~A. Graur, Y.~H. Zhang, J.~M. Reese, {Comparative study of
  the Boltzmann and McCormack equations for Couette and Fourier flows of binary
  gaseous mixtures}, Int. J. Heat Mass Flow 96 (2016) 29--41.

\bibitem{Lei2015Enskog}
L.~Wu, Y.~Zhang, J.~M. Reese, {Fast spectral solution of the generalized Enskog
  equation for dense gases}, J. Comput. Phys. 303 (2015) 66--79.

\bibitem{WuJFM2016}
L.~Wu, H.~H. Liu, J.~M. Reese, Y.~H. Zhang, Non-equilibrium dynamics of dense
  gas under tight confinement, J. Fluid Mech. 794 (2016) 252--266.

\bibitem{Filbet2012a}
F.~Filbet, J.~Hu, S.~Jin, A numerical scheme for the quantum {B}oltzmann
  equation with stiff collision terms, Math. Model. Num. Anal. {46} ({2012})
  {443--463}.

\bibitem{Hu2012}
J.~Hu, L.~Ying, {A fast spectral algorithm for the quantum Boltzmann collision
  operator}, Commun. Math. Sci. {10} ({2012}) {989--999}.

\bibitem{Somme2011}
A.~Sommer, M.~Ku, G.~Roati, M.~W. Zwierlein, {Universal spin transport in a
  strongly interacting Fermi gas}, Nature 472 (2011) 201--204.

\bibitem{Sommer2011b}
A.~Sommer, M.~Ku, M.~W. Zwierlein, {Spin transport in polaronic and superfluid
  Fermi gases}, New J. Phys. 13 (2011) 055009.

\bibitem{Koschorreck2013}
M.~Koschorreck, D.~Pertot, E.~Vogt, M.~K\"{o}hl, {Universal spin dynamics in
  two-dimensional Fermi gas}, Nat. Phys. 9 (2013) 405--409.

\bibitem{Filbet2006}
F.~Filbet, C.~Mouhot, L.~Pareschi, {Solving the Boltzmann equation in NlogN},
  SIAM J. Sci. Comput. {28} ({2006}) 1029--1053.

\bibitem{WuJFM2017Boundary}
L.~Wu, H.~Struchtrup, {Assessment and development of the gas kinetic boundary
  condition for the Boltzmann equation}, J. Fluid Mech. 823 (2017) 511--537.

\bibitem{Pareschi2000}
L.~Pareschi, G.~Russo, {Numerical solution of the Boltzmann equation I:
  Spectrally accurate approximation of the collision operator}, SIAM J. Numer.
  Anal. {37} ({2000}) {1217--1245}.

\bibitem{bruun_2012}
G.~M. Bruun, {Shear viscosity and spin-diffusion coefficient of a
  two-dimensional Fermi gas}, Phys. Rev. A 85 (2012) 013636.

\bibitem{arxiv_sch}
T.~Sch\"{a}fer, Shear viscosity and damping of collective modes in a
  two-dimensional {Fermi} gas, Phys. Rev. A 85 (2012) 033623.

\bibitem{Watabe2010}
S.~Watabe, A.~Osawa, T.~Nikuni, Zero and first sound in normal {Fermi} systems,
  J. Low Temp. Phys. {158} ({2010}) {773--805}.

\bibitem{Smith_book}
H.~Smith, H.~H. Jensen, {Transport Phenomena}, Oxford University Press, 1989.

\bibitem{Nikuni1998}
T.~Nikuni, A.~Griffin, {Hydrodynamic damping in trapped Bose gases}, J. Low
  Temp. Phys. {111} ({1998}) {793--814}.

\bibitem{viscosity_entropy}
P.~K. Kovtun, D.~T. Son, A.~O. Starinets, Viscosity in strongly interacting
  quantum field theories from black hole physics, Phys. Rev. Lett. 94 (2005)
  111601.

\bibitem{Song2010Transport}
B.~Song, X.~Wang, J.~Wu, Z.~Liu, Prediction of transport properties of pure
  noble gases and some of their binary mixtures by ab initio calculations,
  Fluid Phase Equilibria 290 (2010) 55--62.

\bibitem{Sharipov2017JCP}
F.~Sharipov, V.~J. Benites, Transport coefficients of helium-neon mixtures at
  low density computed from ab initio potentials, J. Chem. Phys. 147 (2017)
  224302.

\bibitem{Wu2012modelling}
L.~Wu, J.~P. Meng, Y.~H. Zhang, Kinetic modelling of the quantum gases in the
  normal phase, Proc. R. Soc. A 468 (2012) 1799--1823.

\bibitem{Yang2013ES}
J.~Y. Yang, C.~Y. Yan, M.~Diaz, J.~C. Huang, Z.~H. Li, H.~X. Zhang, Numerical
  solutions of ideal quantum gas dynamical flows governed by semiclassical
  ellipsoidal-statistical distribution, Proc. R. Soc. A 470 (2013) 20130413.

\bibitem{Diaz2016JCP}
M.~A. Diaz, J.~Y. Yang, An efficient direct solver for rarefied gas flows with
  arbitrary statistics, J. Comput. Phys. 305 (2016) 127--149.

\bibitem{Changliu2016}
C.~Liu, K.~Xu, Q.~H. Sun, Q.~D. Cai, {A unified gas-kinetic scheme for
  continuum and rarefied flows IV: Full Boltzmann and model equations}, J.
  Comput. Phys. 314 (2016) 305--340.

\bibitem{XiaoTB2017JCP}
T.~B. Xiao, Q.~D. Cai, K.~Xu, A well-balanced unified gas-kinetic scheme for
  multiscale flow transport under gravitational field, J. Comput. Phys. 332
  (2017) 475--491.

\end{thebibliography}

\end{document}